\documentclass[aps,prb,groupedaddress,superscriptaddress,twocolumn]{revtex4-2}
\usepackage{amsmath,graphicx}
\usepackage{float}
\usepackage{subfigure}
\usepackage{color}
\usepackage{braket}
\usepackage{mathrsfs}
\usepackage{appendix}

\begin{document}
\title{Phase diagrams and edge-state transitions in graphene with spin-orbit coupling and magnetic and pseudomagnetic fields}
\author{Yu-Chen Zhuang}
\affiliation{International Center for Quantum Materials, School of Physics, Peking University, Beijing 100871, China}
\affiliation{CAS Center for Excellence in Topological Quantum Computation, University of Chinese Academy of Sciences, Beijing 100190, China}

\author{Qing-Feng Sun}
\email[]{sunqf@pku.edu.cn}
\affiliation{International Center for Quantum Materials, School of Physics, Peking University, Beijing 100871, China}
\affiliation{CAS Center for Excellence in Topological Quantum Computation, University of Chinese Academy of Sciences, Beijing 100190, China}
\affiliation{Beijing Academy of Quantum Information Sciences, West Bld.\#3, No.10 Xibeiwang East Rd., Haidian District, Beijing 100193, China}

\date{\today}

\begin{abstract}
The quantum Hall (QH) effect, the quantum spin Hall (QSH) effect
and the quantum valley Hall (QVH) effect are
three peculiar topological insulating phases in graphene. They are characterized by three different types of edge states.
These three effects are caused by the external magnetic field,
the intrinsic spin-orbit coupling (SOC) and
the strain-induced pseudomagnetic field, respectively.
Here we theoretically study phase diagrams when these effects coexist
and analyze how the edge states evolve between the three.
We find the real magnetic field and the pseudomagnetic field will
compete above the SOC energy gap
while the QSH effect is almost unaffected within the SOC energy gap.
The edge states transition from the QH effect or the QVH effect
to the QSH effect directly relies on the arrangement of the zeroth Landau levels.
Using edge states transitions, we raise a device similar to a spin field effect transistor (spin-FET) and also design a spintronics multiple-way switch.
\end{abstract}

\maketitle	
\section{\label{sec1} Introduction}
Graphene, a two-dimensional crystal with a hexagonal honeycomb lattice structure, has long received extensive attention due to its unique physical properties \cite{Geim1,Geim2}. It has a special energy band structure where the conduction and valence bands touch at two inequivalent points $K$ and $K'$. These two points are often dubbed as valleys, corresponding to a pseudospin degree of freedom \cite{Castro1}. The low-energy excitations around these valleys or so-called Dirac points are massless and chiral Dirac fermions which could carry a nontrivial topological Berry phase \cite{Novoselov1, Mikitik,refa1,refa2}. Based on this, graphene can serve as a platform to realize many novel topological phases.

One typical topological phase in graphene is the  ``relativistic'' quantum Hall (QH) effect when the graphene is subjected to a relatively high magnetic field \cite{Zhang, Goerbig}. Different from the ``conventional" QH effect in the two-dimensional electron gas, QH states in graphene display a series of Hall plateaus at filling factors $\nu=4(n+\frac{1}{2})$ with $n=0, \pm1, \pm2,...$ (4 comes from valley and spin degeneracy) \cite{Castro1,Castro2,Goerbig,Peres1}. These QH plateaus can be observed even at room temperature due to large cyclotron energies \cite{Novoselov2}. When the Fermi energy lies between Landau levels (LLs), although the bulk is insulating, the Hall current is carried by the spin-degenerate Hall edge states at sample boundaries, with a definite chirality \cite{Castro2,Brey,refa3,Chen2}.

Another exotic topological phase in graphene is the quantum spin Hall (QSH) effect \cite{Hasan,Qi}. It was proposed in the Kane-Mele model in which the spin-orbit coupling (SOC) can open up a topological energy gap at two inequivalent valleys for two sets of spins \cite{Kane1, Kane2}. The QSH effect in graphene can be simply regarded as a spin version of Haldane model if only the intrinsic SOC is considered \cite{Haldane, Kane1, Hasan}. Within the SOC energy gap, the QSH effect is characterized by helical edge states where two opposite spin states propagate towards two opposite directions \cite{Hasan,Qi}. Due to the protection of time-reversal symmetry, helical edge states are robust against nonmagnetic scattering and dephasing \cite{Kane1,Lu,Jiang}. Although the QSH effect in graphene has so far not been observed experimentally because the energy gap opened by SOC is so small \cite{Min,Yao}, some theories and experiments have pointed out that the introduction of adatoms \cite{Weeks,Jiang2,Balakrishnan} or the proximity effect \cite{Avsar1, Wang} could considerably improve the strength of SOC for several orders of magnitude \cite{Avsar2}. These approaches make it possible to realize the QSH effect in graphene.

Since graphene is a one-atom sheet and amenable to external influences \cite{Lee}, mechanical deformation (i.e. strain engineering) becomes an effective means to tailor its electronic properties \cite{Pereira,Amorim,Peres2}. Strain modifies the hopping parameters between graphene lattices and introduces a gauge vector near Dirac points \cite{Guinea1}. One of the most outstanding phenomena is that an inhomogeneous strain field can induce a pseudomagnetic field \cite{Ramezani}. Unlike real magnetic fields, pseudomagnetic fields maintain time-reversal symmetry, thus are opposite at two valleys. They can lead to a quantum phase of ``pseudo-quantum Hall effect" or ``quantum valley Hall (QVH) effect" \cite{Guinea2,Low}. The QVH effect could be roughly regarded as a counterpart of the QSH effect, while the edge states of LLs from opposite valleys propagate towards opposite directions at the same edge of the graphene \cite{Low,Wu}. Extremely strong pseudomagnetic fields and LL signals are observed in the strained graphene in many experiments \cite{Levy,Jiang3,Mao,Meng,Georgi,Jia}. A theoretical scheme also proposed that a large, nearly uniform and tunable pseudomagnetic field could be generated by an uniaxial stretch \cite{Zhu}, which provides a promising way to implement the QVH effect in graphene.

A question naturally arises that what will happen
if three aforementioned typical effects are combined in graphene.
Several literatures discussed the influence of magnetic fields
or pseudomagnetic fields on the QSH effect in graphene respectively \cite{Shevtsov2,Beugeling,Martino,He}. However, the edge states transitions, the coexistence and the competition among these three phases still remain illusive.

Our work is aimed to shed light on these questions. We find that the SOC energy gap of the QSH effect can keep whether in a real magnetic field or a pseudomagnetic field. Outside the SOC energy gap, the real magnetic field and pseudomagnetic field will compete with each other, affecting the LLs for different valleys. We also find that the transitions of the edge states
from the QH effect or the QVH effect to the QSH effect
are quite different
and directly relate to their distinctions on the arrangements of the zeroth LLs. Based on above discoveries, we construct a valuable spintronics device to regulate the spin polarization of edge states and achieve a multiple-way switch to control the spin current to flow into different terminals.

\begin{figure}[ht]
	\includegraphics[width=1\columnwidth]{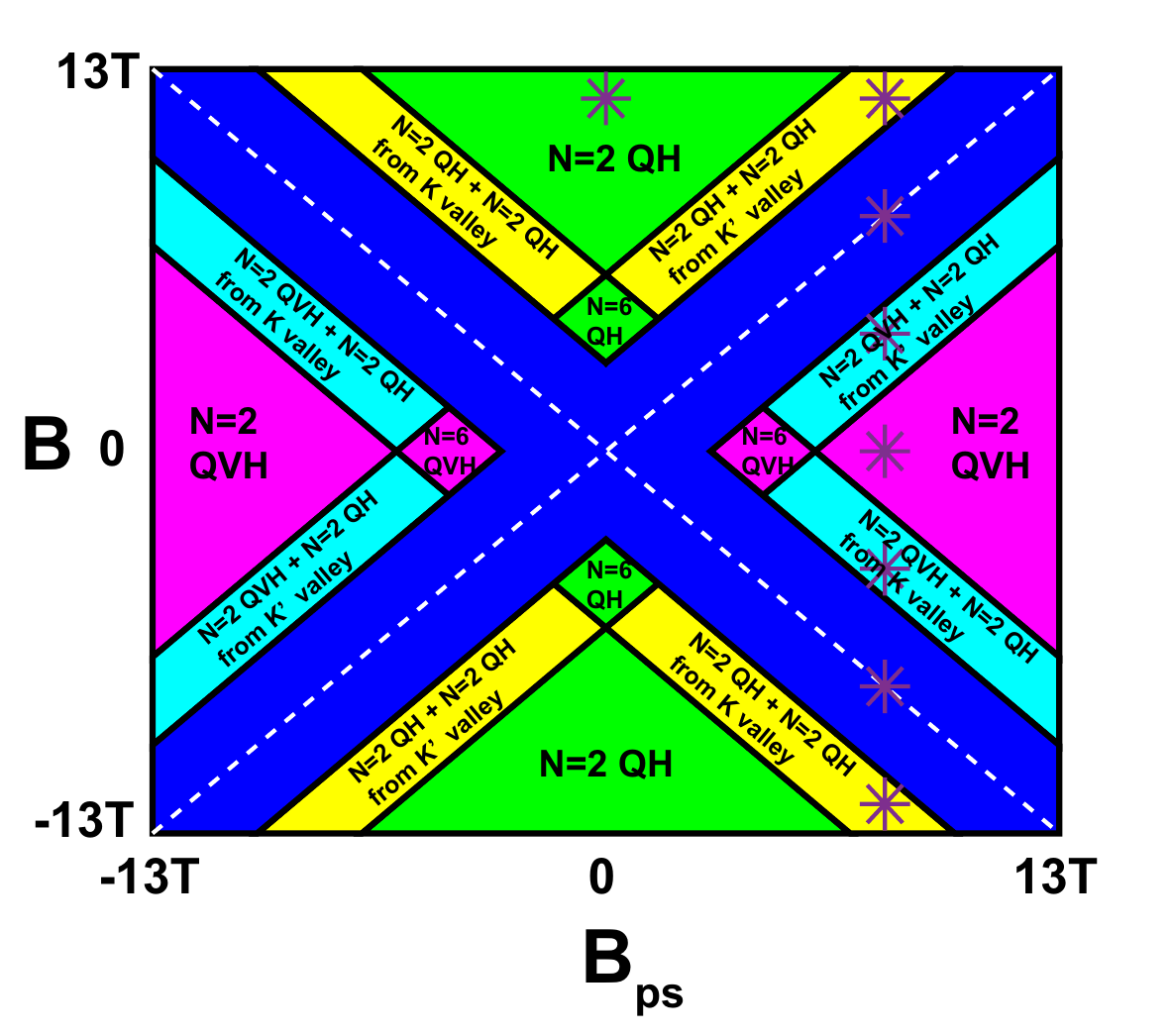}
	\centering 
	\caption{The phase diagram for the graphene where a real magnetic field $B$ and a pseudomagnetic field $B_{ps}$ coexist with a fixed Fermi level $E_{f}$. The colored patches denote different QH/QVH phases (other higher filling factor phases are all denoted by dark blue). White dotted lines denote the boundaries $\vert B\vert =\vert B_{ps} \vert$. $E_{f}=\sqrt{2e \hbar v_{f}^{2}*6}\approx 0.0805eV$.
	}
	\label{FIG1}
\end{figure}

\begin{table*}[t]
	\begin{center}
		\caption{The table for LL eigenvalues and the zeroth LL eigenstates for different valleys $K/K'$ and spin-up $\uparrow$ and spin-down $\downarrow$ components under a real magnetic field $B$ or pseudomagnetic field $B_{ps}$ along with the SOC energy gap $\Delta_{so}$. The spinors of wavefunctions $\Psi$ represent A and B sublattices. The eigenstates $\vert 0\rangle $ and $\vert \widetilde{0}\rangle $ refer to the ground states for the harmonic oscillator $\mathbf{a}^{\dagger}\mathbf{a}$ with $\mathbf{a}=\frac{l_{B}}{\sqrt{2}}(k_{x}-y/l_{B}^{2}-\partial_{y})$ and $\mathbf{\widetilde{a}}^{\dagger}\mathbf{\widetilde{a}}$ with $\mathbf{\widetilde{a}}=\frac{l_{B}}{\sqrt{2}}(k_{x}+y/l_{B}^{2}+\partial_{y})$\cite{Lee2}. Here $\epsilon_{n}=E_{n}/\frac{\hbar v_{f}}{l_{B}}$ where $E_{n}$ is the energy for the n-th LL $(n=0,1,2, ...)$, $\tilde{\Delta}_{so}=\Delta_{so}/\frac{\hbar v_{f}}{l_{B}}$, $l_{B}=\sqrt{\frac{\hbar}{eB}}$ or $\sqrt{\frac{\hbar}{eB_{ps}}}$ is the magnetic length. }

\begin{tabular}{|c|c|c|c|c|}
\hline
Types &spin $\uparrow$ for $K$ valley &spin $\downarrow$ for $K$ valley&spin $\uparrow$ for $K'$ valley & spin $\downarrow$ for $K'$ valley\\ \hline The n-th LLs
$\epsilon_{n \neq 0}$ for $B$ &$\epsilon_{\pm n}\pm \sqrt{2 n +\tilde{\Delta} _{so}^{2} }$ &$\epsilon_{\pm n}=\pm \sqrt{2n +\tilde{\Delta} _{so}^{2} }$&$\epsilon_{\pm n}=\pm \sqrt{2n +\tilde{\Delta}_{so}^{2} }$&$\epsilon_{\pm n}=\pm \sqrt{2n+\tilde{\Delta} _{so}^{2} }$\\ \hline
The zeroth LLs $\epsilon_{0}$ for $B$ &$\epsilon_{0}=-\tilde{\Delta}_{so}$ &$\epsilon_{0}=\tilde{\Delta}_{so}$&$\epsilon_{0}=-\tilde{\Delta}_{so}$ & $\epsilon_{0}=\tilde{\Delta}_{so}$\\ \hline
The zeroth LL $\Psi_0$ for $B$ & $\binom{0}{\vert 0\rangle }$ &$\binom{0}{\vert 0\rangle}$&$\binom{\vert 0\rangle}{0}$&$\binom{\vert 0\rangle }{ 0 }$\\ \hline The n-th LLs $\epsilon_{n \neq 0}$ for $B_{ps}$ &$\epsilon_{\pm n}=\pm \sqrt{2n+\tilde{\Delta} _{so}^{2} }$ &$\epsilon_{\pm n}=\pm \sqrt{2n+\tilde{\Delta} _{so}^{2} }$&$\epsilon_{\pm n}=\pm \sqrt{2n+\tilde{\Delta} _{so}^{2} }$&$\epsilon_{\pm n}=\pm \sqrt{2n+\tilde{\Delta} _{so}^{2} }$\\ \hline
The zeroth LLs $\epsilon_{0}$ for $B_{ps}$ &$\epsilon_{0}=-\tilde{\Delta}_{so}$ &$\epsilon_{0}=\tilde{\Delta}_{so}$&$\epsilon_{0}=\tilde{\Delta}_{so}$ & $\epsilon_{0}=-\tilde{\Delta}_{so}$\\ \hline
The zeroth LL $\Psi_0$ for $B_{ps}$ & $\binom{0}{\vert 0\rangle }$ &$\binom{0}{\vert 0\rangle}$&$\binom{0}{\vert \widetilde{0} \rangle }$&$\binom{0}{\vert \widetilde{0} \rangle}$ \\
\hline
\end{tabular}
\label{Table1}
\end{center}
\end{table*}

The architecture of our article adopts a progressive layer-by-layer approach. We first examine the relationship between each of the two phases separately, and finally investigate the comprehensive effect of all three phases. The rest of this paper is organized as follows.
In Sec II, we present our model, Hamiltonian and solutions of LLs in graphene with SOC under a real magnetic field or a pseudomagnetic field.
In Sec III, ignoring SOC, we study the competition between the real magnetic field and the strain-induced pseudomagnetic field in a zigzag graphene nanoribbon. In Sec IV, we ignore the pseudomagnetic field and study the influence of a real magnetic field on the QSH effect. In Sec V, we parallelly analyze the influence of a pseudomagnetic field on the QSH effect. The results are also compared with those in Sec IV. Based on these, we further propose a possible spin FET-like device to adjust the spin polarization of the edge states. In Sec VI, we combine the real magnetic field, the pseudomagnetic field and SOC in graphene together. The relationship of their competition and coexistence is analyzed in detail. Moreover, we design a spintronics multiple-way switching device. In Sec VII, we give our conclusions. Some discussions about the effect of strain on SOC strength and the effect of Rashba SOC on the LLs and topological phases are put in the Appendix A and Appendix B.

\section{\label{sec2} The model and Hamiltonian}

We consider an infinite zigzag graphene nanoribbon with a periodic direction along $x$ and lattice number $N_{y}$ along y direction. Considering strain-modulated hopping parameters, the tight-binding Hamiltonian based on the Kane-Mele model with a real magnetic field can be written as \cite{Kane1,Shevtsov2}
\begin{equation}
	H=-\sum_{\langle ij \rangle ,\alpha}t_{ij}e^{i\phi_{ij}}c^{\dagger}_{i\alpha}c_{j\alpha}+\sum_{\langle \langle ij\rangle \rangle ,\alpha,\beta}i\lambda_{so,ij}e^{i\phi_{ij}}\upsilon_{ij}
c^{\dagger}_{i\alpha}s^{z}_{\alpha\beta}c_{j\beta}.
	\label{Eq1}
\end{equation}
Here indices $(i,j) $ denote lattice sites, $\alpha$ and $\beta=\uparrow, \downarrow$ denote spin indices, and $s^z$ is the Pauli matrix.
$\langle \rangle $ and $\langle \langle \rangle \rangle $ refer to nearest-neighbor and next-nearest-neighbor bonds. The first term is about the nearest-neighbor hoppings between lattices with the hopping energy $t_{ij}$.
The second term is about the intrinsic SOC with the SOC strength $\lambda_{so,ij}$. $\upsilon_{ij}=\pm 1$ depends on the orientation of two nearest-neighbor bonds as the electron transverses from lattice $j$ to $i$. $\phi_{ij}=\frac{2\pi}{\phi_{0}}\int_{\mathbf{r}_{j}}^{\mathbf{r}_{i}} \vec{A} \cdot \,d\vec{r} $ is Peierls phase from the real magnetic field with
the flux quantum $\phi_{0}$ and the vector potential $\vec{A}$.
To include the strain effect, the hopping parameters between strained lattices can be modulated as \cite{Pereira2}
\begin{equation}
	t_{ij}=t_{0}e^{-\gamma( d_{ij}/a-1) }.
	\label{Eq2}
\end{equation}
$t_{0}=2.8eV$ is the unstrained nearest-neighbor hopping energy and $d_{ij}$ is the distance between the strained lattices $i$ and $j$. $a=0.142nm$ is the unstrained carbon-carbon bond distance. $\gamma$ is the Gruneisen parameter of graphene and often taken as $2-3.37$ \cite{Ramezani,Lantagne}.
We take $\gamma=3$ in the following numerical calculations but actually it does not matter.
In principle, the modulations of the intrinsic SOC strength by strain
should also be considered. However, the intrinsic SOC strength only brings about
a SOC energy gap near the Dirac points. The higher order corrections from strain are too small to affect the results [See details in the Appendix A].
So in the following calculations, we set $\lambda_{so,ij}$ in Equation~(\ref{Eq1}) as a constant $\lambda_{so}$
not depending on the lattice positions $(i,j)$. In addition, the effect of Rashba SOC is ignored here, which may break the QSH phase and result in spin splitting (See details in the Appendix B).

Ignoring the magnetic field and intrinsic SOC, we first review how strain field brings about a pseudomagnetic field under the continuum elasticity frame.
$d_{ij}$ is a smooth function of spatial coordinates $( x,y ) $ and can be expressed as \cite{Ramezani}
\begin{equation}
\vec{d}_{ij}=( I+\bar{\epsilon}) \vec{\delta}_{ij}.
\label{Eq3}
\end{equation}
$\bar{\epsilon}_{\imath\jmath}=\frac{1}{2}
(\partial_{\imath}u_{\jmath}+\partial_{\jmath} u_{\imath} )$
with $\imath/\jmath =x,y$
is the strain tensor defined from an in-plane displacement field $\vec{u}=( u_{x}, u_{y} )$. $\vec{\delta}_{ij}$ refers to the unstrained nearest-neighbor bond vector. There are only three $\vec{\delta}_{ij}$ in the periodic lattices of graphene with: $\vec{\delta}_{1}=(0,a)$, $\vec{\delta}_{2}=(-\frac{\sqrt{3}}{2}a,-\frac{1}{2}a)$ and $\vec{\delta}_{3}=(\frac{\sqrt{3}}{2}a,-\frac{1}{2}a)$.
As the strain is small, the hopping parameters corresponding to these nearest-neighbor bonds $\vec{\delta}_{n}$ in Equation~(\ref{Eq2}) can be linearly expanded into $t_{n} \approx t_{0}( 1+\delta t_{n} )$ with \cite{Lantagne}
\begin{equation}
	\delta t_{n}= -\frac{\gamma}{a^{2}}\vec{\delta_{n}}\cdot \bar{\epsilon} \cdot \vec{\delta_{n}}.
	\label{Eq4}
\end{equation}
Using the Fourier transformation, the strain Hamiltonian in Equation~(\ref{Eq1}) can be written in the $k$ space (neglecting the Peierls phase, spin and intrinsic SOC): \cite{Pereira}
\begin{equation}
	H=-\sum_{n,k}t_{n}e^{i\vec{k}\cdot \vec{d}_{n}} c^{\dagger}_{A,k}c_{B,k}+H.c.
	\label{Eq5}
\end{equation}
$(A,B)$ refer to sublattice indices. $\vec{d}_{n}$ is the strained bond vector of $\vec{\delta}_{n}$ as Equation~(\ref{Eq3}) indicates.  Putting Equations~(\ref{Eq3},\ref{Eq4}) into Equation~(\ref{Eq5}) and focusing on the low-energy Hamiltonian near the Dirac point $K$, the first order modification from strain generates a pseudovector potential $\mathbf{A^{ps}}$: \cite{Pereira}
\begin{equation}
	\mathbf{A^{ps}}=A_{x}^{ps}-iA_{y}^{ps}=-\frac{1}{e v_{f}}\sum_{n}t_{0}\delta t_{n}e^{i\vec{K}\cdot \vec{\delta}_{n}}
	\label{Eq6}
\end{equation}
Equation~(\ref{Eq6}) connects hopping energy modulations and the strain-induced pseudovector potential:
\begin{equation}
    \begin{split}
	A_{x}^{ps}(x,y)&=-\frac{t_{0}}{2e v_{f}}(2\delta t_{1}-\delta t_{2}-\delta t_{3}) \\
	A_{y}^{ps}(x,y)&=-\frac{\sqrt{3} t_{0}}{2e v_{f}}(\delta t_{2}-\delta t_{3}).
    \end{split}
	\label{Eq7}
\end{equation}
$v_{f}=\frac{3t_{0}a}{2\hbar}\approx 9*10^{5}m/s$ is the Fermi velocity and $e$ is the electron charge. With the help of Eq.~(\ref{Eq4}), the relation between the strain tensor and strain-induced pseudovector is: \cite{Settnes}
\begin{equation}
	\begin{split}
	A_{x}^{ps}(x,y)&=-\frac{\phi_{0}\gamma}{4\pi a}(\bar{\epsilon}_{xx}-\bar{\epsilon}_{yy}) \\
	A_{y}^{ps}(x,y)&=-\frac{\phi_{0}\gamma}{4\pi a}(-2\bar{\epsilon}_{xy}).
	\end{split}
	\label{Eq8}
\end{equation}
$\phi_{0}=h/e$ is the flux quantum.
Since time reversal symmetry is unbroken, the pseudovector $\mathbf{A^{ps}}$ should be opposite at two inequivalent valleys $(K,K')$. In our following studies, an uniaxial inhomogeneous strain field along the armchair direction (y direction) will be considered. This means $\bar{\epsilon}_{xx}=\bar{\epsilon}_{xy}=0$. $\bar{\epsilon}_{yy}$ is a linear function of position $y$. This configuration could generate a large uniform pseudomagnetic field $B_{ps}$ spreading over the graphene nanoribbon.

When applying a (pseudo)magnetic field on graphene with the intrinsic SOC (Kane-Mele model), a gauge vector is introduced to replace the $\vec{p}$ as $\vec{p}+e\vec{A}$.  The low-energy Hamiltonian for the Kane-Mele model under a magnetic field and
pseudomagnetic field is: \cite{Martino}
\begin{equation}
	\begin{split}
	H=&v_{f}[ \sigma_{x}\tau_{z}(p_{x}+eA_{x}+e\tau_{z}A_{x}^{ps})+\sigma_{y}(p_{y}+eA_{y}+e\tau_{z}A_{y}^{ps}) ] \\
	&+\Delta_{so}\sigma_{z}s_{z}\tau_{z}
	\end{split}
	\label{Eq9}
\end{equation}
where $\sigma$ denotes $(A,B)$ sublattice, $\tau$ denotes valley $(K,K')$ and $s$ denotes spin (up $\uparrow$, down $\downarrow$). $A_{x,y}$ is the vector potential from the real magnetic field $B$ and $A_{x,y}^{ps}$ is the vector potential from the pseudomagnetic field $B_{ps}$. $\Delta_{so}=3\sqrt{3} \lambda_{so}$ is the energy gap opened by SOC $\lambda_{so}$. The eigenvalues and eigenstates for Eq.~(\ref{Eq9}) are easy to be solved \cite{Martino}. Here we focus on the cases where only $B$ exists or $B_{ps}$ exists along with the SOC gap $\Delta_{so}$. All the results are summarized in Table~\ref{Table1}. For the $B$ case, the energy of non-zero n-th LLs is spin and valley-degenerate ($E_{\pm n}=\pm \sqrt{2e\hbar v_{f}^{2}Bn+\Delta_{so}^{2}}, n=1,2,...$). However, spin degeneracy is broken on the zeroth LLs \cite{Martino,Shevtsov2}. The energy of the zeroth LLs for the spin-up component is $E_{0}=-\Delta_{so}$ at both valleys while the energy of zeroth LLs for the spin-down component is $E_{0}=\Delta_{so}$ at both valleys [see Table~\ref{Table1}]. Additionally, the zeroth LL wavefunctions under $B$ exhibit distinctive sublattice polarizations at different valleys (e.g. B sublattice polarization at the $K$ valley and A sublattice polarization at the $K'$ valley, as shown in Table~\ref{Table1}). For the $B_{ps}$ case, the results for $K'$ valley change as  $B_{ps}$ should reverse sign at $K'$ valley. The energy of non-zero n-th LLs is still spin and valley-degenerate ($E_{\pm n}=\pm \sqrt{2e\hbar v_{f}^{2}B_{ps}n+\Delta_{so}^{2}},n=1,2,...$). But for the zeroth LLs at the $K'$ valley, the energy of the spin-up component is $E_{0}=\Delta_{so}$ and of the spin-down component is $E_{0}=-\Delta_{so}$. In general, both spin and valley degeneracy of the zeroth LLs are broken under the pseudomagnetic field and intrinsic SOC, but they exhibit a spin-valley locking configuration, which guarantees the time-reversal symmetry ($E^{K}_{0,\uparrow}=E^{K'}_{0,\downarrow}=-\Delta_{so}$ and
$E^{K}_{0,\downarrow}=E^{K'}_{0,\uparrow}=\Delta_{so}$). There also exhibits the same sublattice polarization for the zeroth LL wavefunctions at both valleys (e.g. B sublattice polarization at both valleys, as shown in Table~\ref{Table1}), which is a typical characteristic of pseudomagnetic fields \cite{Settnes,Mao}. Actually, these results are still valid when $B$ and $B_{ps}$ coexist except that the total magnetic field for $K$ valley is $B+B_{ps}$ and for $K'$ valley is $B-B_{ps}$. In addition, the distinctiveness of the zeroth LL can be reflected from the LL spectrum versus (pseudo)magnetic field $B$ $(B_{ps})$. For the non-zero n-th LLs, they depend on the (pseudo)magnetic field $B$ $(B_{ps})$ by relation $E_{\pm n}=\pm \sqrt{2e\hbar v_{f}^{2}Bn+\Delta_{so}^{2}}$ $(E_{\pm n}=\pm \sqrt{2e\hbar v_{f}^{2}B_{ps}n+\Delta_{so}^{2}})$. For the degeneracy-broken zeroth LLs, their energy is stuck on $\pm \Delta_{so}$ and does not change with the (pseudo)magnetic field $B$ ($B_{ps}$).

\section{\label{sec3} The coexistence of a real magnetic and a pseudomagnetic field in graphene nanoribbon}

In this section, the coexistence of the real magnetic and pseudomagnetic field on graphene with vanishing SOC $(\lambda_{so}= 0)$ will be investigated. As Eq.~(\ref{Eq9}) suggests, both of them could induce LLs with energy $E_{\pm n}=\pm \sqrt{2 e \hbar \vert \mathcal{B}  \vert  v_{f}^{2}n}$ ($n=0,1,2,..$). $\vert \mathcal{B}  \vert$ is the magnitude of the total magnetic field. The main difference between the real magnetic field $B$ and pseudomagnetic field $B_{ps}$ is that the former is the same for two valleys, but the latter is opposite. When both of them exist, the total magnetic fields for two valleys become imbalanced, resulting a LL valley splitting that has been observed in the experiments \cite{Li1,Li2}. As table~\ref{Table1} indicates ($\Delta_{so}=0$), the n-th LLs $(n=0,1,2,..)$ at $K$ valley and $K'$ valley are \cite{Roy1, Roy2}
\begin{equation}
	\begin{split}
	E_{\pm n}^{K}&=\pm \sqrt{2 e \hbar \vert B+B_{ps} \vert  v_{f}^{2}n}  \\
	E_{\pm n}^{K'}&=\pm \sqrt{2 e \hbar \vert B-B_{ps} \vert  v_{f}^{2}n}.
	\end{split}
\label{Eq10}
\end{equation}
Here can be firstly divided into two regions \cite{Roy2}: (1) $B$-dominated region ($\vert B\vert > \vert B_{ps} \vert$). The signs of total magnetic fields for two valleys are same. All edge states have the same chirality. A typical representative is the QH phase. (2) $B_{ps}$-dominated region ($\vert B\vert < \vert B_{ps} \vert$). The signs of total magnetic fields for two valleys are opposite. Edge states from two valleys thus have the opposite chirality. A typical representative is the QVH phase. Moreover, in each case, if the Fermi level just lies between $E_{n}^{K}$ and $E_{n}^{K'}$, the valley degeneracy of LLs will be broken. More LLs from one valley will be crossed and contribute more edge states. At this time, the system will be further classified as valley-imbalanced QH or QVH phases \cite{Roy1,Roy2,Low}.

To illustrate above process in detail, we exhibit the phase diagram where a real magnetic field $B$ and pseudomagnetic field $B_{ps}$ coexist in Fig.~\ref{FIG1}. In this diagram, the Fermi level is fixed at $E_{f}=\sqrt{2e \hbar v_{f}^{2}*6}\approx 0.0805eV$. We only focus on those low filling factor phases since $E_{f}$ is close to the charge neutral point (other higher filling factor phases with more occupied LLs below the Fermi level are denoted by dark blue). To label these QH or QVH phases, the filling factor $N$, i.e. the number of occupied LLs are labeled. In particular, for those valley-imbalanced phases, we use the `$+$' to denote the filling factor of the additional contribution from one valley.

If $B_{ps}=0$, as $\vert B \vert $ climbs, the energy for each non-zero order LL increases and less LLs are crossed by $E_{f}$. Filling factors $N$ declines (e.g. the $N$=6 QH phase to the $N$=2 QH phase, see green patches in Fig.~\ref{FIG1}).
Similar situations happen if $B=0$ whereas $\vert B_{ps} \vert $ climbs,
except that QH phases are replaced by QVH phases (see magenta patches in Fig.~\ref{FIG1}).
Once $B$ and $B_{ps}$ coexist, phases will become complicated.
For example, starting from the $N$=2 QH phase ($E_f$ only exceeds the zeroth LL) with $B_{ps}=0, B>0$, the energy of the 1st LL from $K'$ valley will decline and tend to be lower than $E_{f}$ as $B_{ps}$ increases. After crossing the phase boundary $E_{f}=\sqrt{2 e \hbar \vert B-B_{ps} \vert  v_{f}^{2}}$, one more spin-degenerate LL from $K'$ valley is occupied. We denote this valley-imbalanced phase as
`$N$=2 QH + $N$=2 QH from $K'$ valley'. Other phase transition processes are similar, with the phase boundaries $E_{f}=\sqrt{2 e \hbar \vert B-B_{ps} \vert  v_{f}^{2}n}$ or $E_{f}=\sqrt{2 e \hbar \vert B+B_{ps} \vert  v_{f}^{2}n}$ with $n=1,2,...$.

For clarity, the topological phases in Fig.~\ref{FIG1} can be further identified by valley Chern numbers $C_{K}/C_{K'}$. We focus on those within the region  $B+B_{ps}>0$ and neglect spin-degeneracy. For $N$=$2,6$ QH phases, $C_{K}=C_{K'}=\frac{1}{2},\frac{3}{2}$. For $N$=$2,6$ QVH phases, $C_{K}=\frac{1}{2},\frac{3}{2}$ while $C_{K'}=-\frac{1}{2},-\frac{3}{2}$. For `$N$=$2$ QH + $N$=$2$ QH from $K'/K$ valley' phase, $C_{K/K'}=\frac{1}{2}$ and $C_{K'/K}=\frac{3}{2}$. For `$N$=$2$ QVH + $N$=$2$ QH from $K'$ valley' phase, $C_{K}=\frac{1}{2}$ and $C_{K'}=-\frac{3}{2}$. For `$N$=$2$ QVH + $N$=$2$ QH from $K$ valley' phase, $C_{K}=\frac{3}{2}$ and $C_{K'}=-\frac{1}{2}$. The valley Chern numbers for topological phases within the region $B+B_{ps}<0$ are just opposite of the above.

\begin{figure*}
	\centering
	\includegraphics[width=0.95\linewidth]{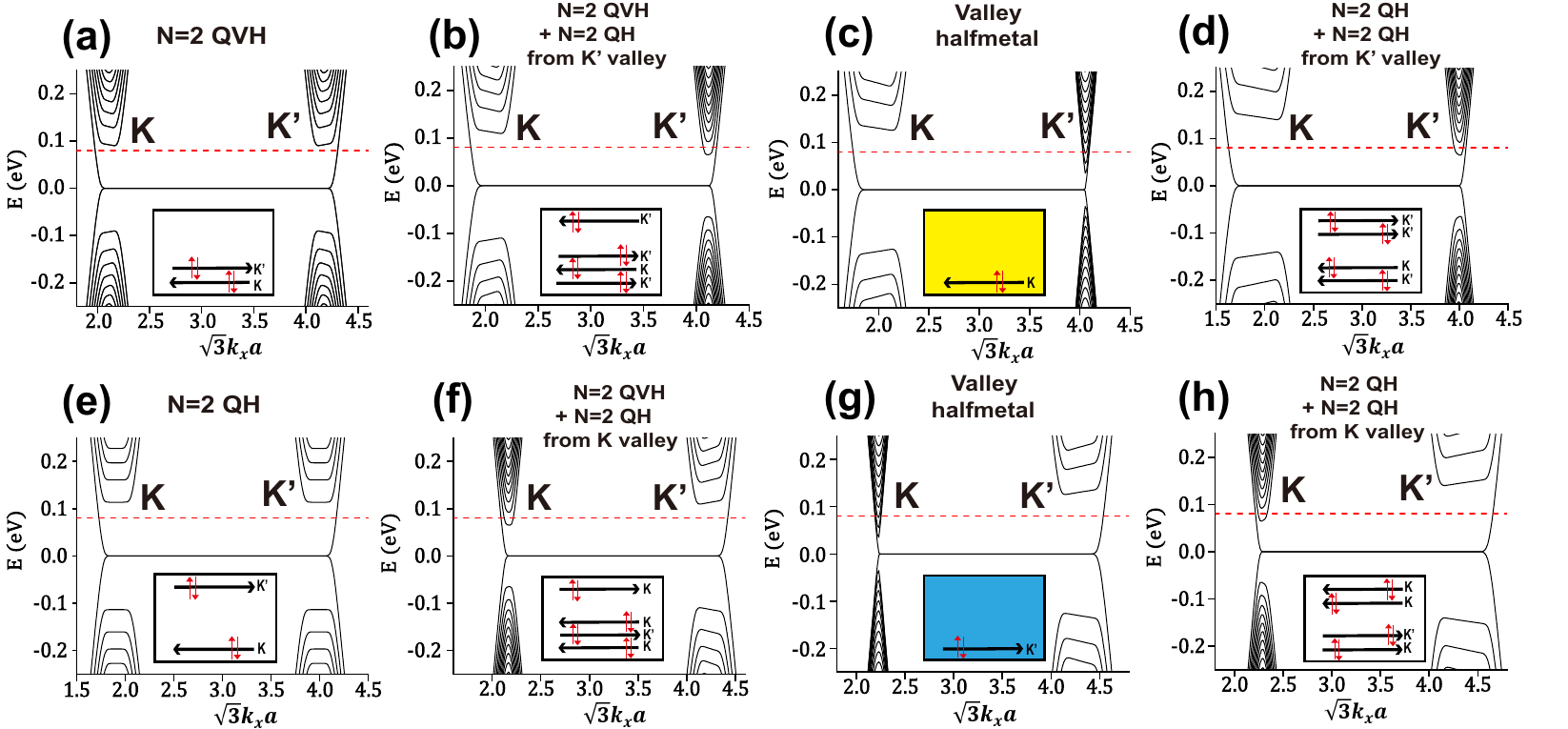}
	\caption{The energy bands for a zigzag graphene nanoribbon without SOC where $B$ and $B_{ps}$ coexist, which correspond to the purple stars in Fig.~\ref{FIG1}. The energy bands here are spin-degenerate. The red dotted lines denote the Fermi level $E_{f}=\sqrt{2e \hbar v_{f}^{2}*6}\approx 0.0805eV$. (a) $B_{ps}=8T,B=0T$. (b)$B_{ps}=8T,B=4T$. (c)$B_{ps}=8T,B=8T$. (d)$B_{ps}=8T,B=12T$. (e)$B_{ps}=0T,B=12T$. (f)$B_{ps}=8T,B=-4T$. (g)$B_{ps}=8T,B=-8T$. (h)$B_{ps}=8T,B=-12T$. The insets within (a-h) schematically show the edge states distributions at Fermi level $E_{f}$ by black solid arrows. The red solid arrows on edge states denote spin-up and spin-down components.
	The yellow and blue colors in (c) and (d) denotes bulk states from $K'$ and K valleys, respectively. $N_{y}=800$.}
	\label{FIG2}
\end{figure*}

These phases in Fig.~\ref{FIG1} can be reflected by the spin-degenerate
energy bands of a zigzag graphene nanoribbon plotted in Fig.~\ref{FIG2}.
The unit cell length along $x$ direction is $\sqrt{3}a$ and the number of lattices is $N_{y}=800$ corresponding to a nanoribbon width $W \approx 90nm$.

The parameters $B$ and $B_{ps}$ follow purple stars in Fig.~\ref{FIG1}. Figures.~\ref{FIG2}(a,b,f) belong to the $B_{ps}$-dominated region and Figs.~\ref{FIG2}(d,e,h) belong to the $B$-dominated region. Figs.~\ref{FIG2}(c,g) are two critical points with $\vert B_{ps} \vert = \vert B \vert $. To feature these phases, the situations of edge states transport at $E_{f}=\sqrt{2e \hbar v_{f}^{2}*6}\approx 0.0805eV$ [denoted by red dotted lines in Fig.~\ref{FIG2}] are schematically present in insets. It is worth noting that for a zigzag graphene nanoribbon, LL edge states have valley polarizations of $K$ valley or $K'$ valley \cite{Tworzydlo,Beenakker,Trifunovic}, which are also marked in the insets.

In Figs.~\ref{FIG2}(a-d), we increase $B$ along the positive direction with a fixed $B_{ps}=8T$. When $B_{ps}=8, B=0$ [Fig.~\ref{FIG2}(a)], the system is in the $N$=2 QVH phase [magenta patch in Fig.~\ref{FIG1}] with dispersive pseudomagnetic field-induced LLs \cite{Lantagne}. A pair of counterpropagating spin-degenerate edge states at $E_{f}$ from two valleys locate at only one edge of the graphene nanoribbon [see inset in Fig.~\ref{FIG2}(a)], unlike two edges in QSH effect \cite{Low,Wu}. The reasons can be understood on two aspects: (i) Due to the magnetic fields at two valleys are opposite, the edge state of the zeroth LL from the K' valley should flip its space location, different from the QH edge states which locate at opposite edges in inset in Fig.~\ref{FIG2}(e). The one-edge valley-polarized channels also correspond to valley Chern number $C_{K}=\frac{1}{2},C_{K'}=-\frac{1}{2}$, and no other edge channels appear at the opposite edge. Actually, this special structure of edge states for the zeroth LL is closely related to chiral anomaly \cite{Lantagne}. (ii) From the symmetry, these counterpropagating modes at the same edge are implied by time-reversal symmetry \cite{Low}. Moreover, inversion symmetry has been broken by strain field in QVH effect, allowing for space asymmetric edge states distributions. Besides, QSH effect and QVH effect also differ in their response to disorders. Different from QSH effect protected by time-reversal symmetry, the QVH effect can exist in the graphene with clean limit or long range disorders, where intervalley scatterings are well suppressed. The short range disorders could couple valleys and destroy the quantized plateaus. The lack of quantization in the presence of the short range disorders
may cause some uncertainty on the verification of QVH effect. However, whether the lower edge [see inset in Fig.~\ref{FIG2}(a)] remains quantized or becomes conductive by short range disorders \cite{Low}, its longitudinal resistance is always quite different from the other insulating edge where no edge channels exist (the upper edge in the inset in Fig.~\ref{FIG2}(a)). The latter tends to exhibit an very large resistance but the former exhibit a finite resistance. This asymmetry resistivity of two opposite edges could give some indications on QVH effect. More importantly, QVH effect with valley polarized edge states can be regarded as a quantum inversion of valley Hall effect \cite{Mak,sui,Lensky}. Similar to valley Hall effect, when applying the longitudinal voltage, opposite valley polarization or orbital momentum polarization will accumulate at opposite edges in QVH effect through valley polarized edge channels (only one edge accumulates valley polarizations for $N=2$ QVH effect) \cite{Bhowal}. This out of plane net magnetization can be detected by Kerr rotation microscopy \cite{Lee3}. 

When $B_{ps}=8T, B=4T$ [Fig.~\ref{FIG2}(b)], the LL energy spacings narrow at $K'$ valley ($B-B_{ps}=-4T$) but widen at $K$ valley ($B+B_{ps}=12T$). One more spin-degenerate LL from $K'$ valley is crossed and additionally contributes two pairs of chiral edge states on the $N$=2 QVH phase. This leads to six edge states appearing at the lower edge while two appearing at the upper edge [see inset in Fig.~\ref{FIG2}(b)]. The system is thus in the ``$N$=2 QVH + $N$=2 QH from $K'$ valley'' phase [cyan patch in Fig.~\ref{FIG1}].
When $B_{ps}=8T, B=8T$ [Fig.~\ref{FIG2}(c)], the gap closes at $K'$ valley
and the energy dispersion astoundingly recovers the Dirac linear dispersion,
while the gap and LLs at $K$ valley hold.
This peculiar phase is a valley halfmetal which was proposed to achieve valley complete polarization transport, since the bulk states all come from one valley [denoted by the yellow color in inset of Fig.~\ref{FIG2}(c)] \cite{Low,Settnes2}.
Once $B_{ps}=8T, B=12T$ [Fig.~\ref{FIG2}(d)], the total magnetic fields are positive at each valley, the gap opens and LLs at $K'$ valley form again.
Both valleys contribute edge states with the same chirality [see inset in Fig.~\ref{FIG2}(d)]. Due to one more spin-degenerate LL from $K'$ valley is crossed ($B-B_{ps}=4T$), we name this phase as ``$N$=2 QH + $N$=2 QH from $K'$ valley'' [yellow patch in Fig.~\ref{FIG1}].
When $B_{ps}=0, B=12T$ [Fig.~\ref{FIG2}(e)], the system is in the $N$=2 QH phase [green patch in Fig.~\ref{FIG1}] and chiral edge states appear at both edges [see inset in Fig.~\ref{FIG2}(e)] \cite{Castro2}.
In Figs.~\ref{FIG2}(f-h), we fix $B_{ps}=8T$
and increase $B$ along the minus direction.
The results are similar to Figs.~\ref{FIG2}(b-d).
When $B_{ps}=8$,$B=-4T$ [Fig.~\ref{FIG2}(f)],
one more spin-degenerate LL from $K$ valley will be crossed by $E_{f}$
and contributes two pairs of chiral edge states on the $N$=2 QVH.
This is the ``$N$=2 QVH + $N$=2 QH from $K$ valley" phase,
similar to the phase in Fig.~\ref{FIG2}(b).
When $B_{ps}=-B=8T$ [Fig.~\ref{FIG2}(g)],
the gap of $K$ valley closes,
the energy band recovers the Dirac linear dispersion,
and the bulk states all come from $K$ valley,
which is also a valley halfmetal.
Once $-B>B_{ps}$, the system is in the $B$-dominated region
and in QH phases.
For example, $B_{ps}=8$ and $B=-12T$ in Fig.~\ref{FIG2}(h),
the system is in the ``$N$=2 QH + $N$=2 QH from $K$ valley" phase.
To conclude, from the $B_{ps}$-dominated region to the $B$-dominated region [e.g. Figs.~\ref{FIG2}(a-d)], the gap of one valley will close and open.
Its edge states will evolve into bulk states and then into edge states again.
Therefore, by simply tuning the real magnetic field in the strained graphene, we can control the location of edge states and achieve a valley polarized transport.

At large real magnetic fields, Zeeman splitting may not be neglected. Since the spin along $z$ direction is a good quantum number in the presence of the intrinsic SOC, Zeeman effect just simply shifts the spin-up and spin-down bands towards opposite directions of the energy axis by $\Delta E_{z}=\frac{1}{2}g\mu_{B}B$. In view that the free electron $g$ factor in graphene is 2 \cite{Volkov}, the Zeeman splitting in graphene under $B=12T$ is roughly $\Delta E_{z} \approx 0.7meV$, which is much smaller than LL spacings. So in Fig.~\ref{FIG2}, the original spin-degenerate bands are slightly split into two spin bands.
Correspondingly, the original topological phases shown in Fig.~\ref{FIG1} and Fig.~\ref{FIG2} will be further divided into spin-polarized QH phases/QVH phases.

\section{\label{sec4} The QSH effect under a real magnetic field.}
In this section, we pay attention to how a real magnetic field affects the QSH phase in graphene. Refs \cite{Shevtsov2, Beugeling,Chen} found the QSH phase could be maintained within the SOC energy gap to a large extent of a real magnetic field, even though the time-reversal symmetry is lack. Moreover, due to the contrast between chiral edge states in QH phases and helical edge states in QSH phases, an ``unhappy" spin has to reverse its direction of propagation when going from one phase to another \cite{Shevtsov2}.

\begin{figure}[t]
	\includegraphics[width=1\columnwidth]{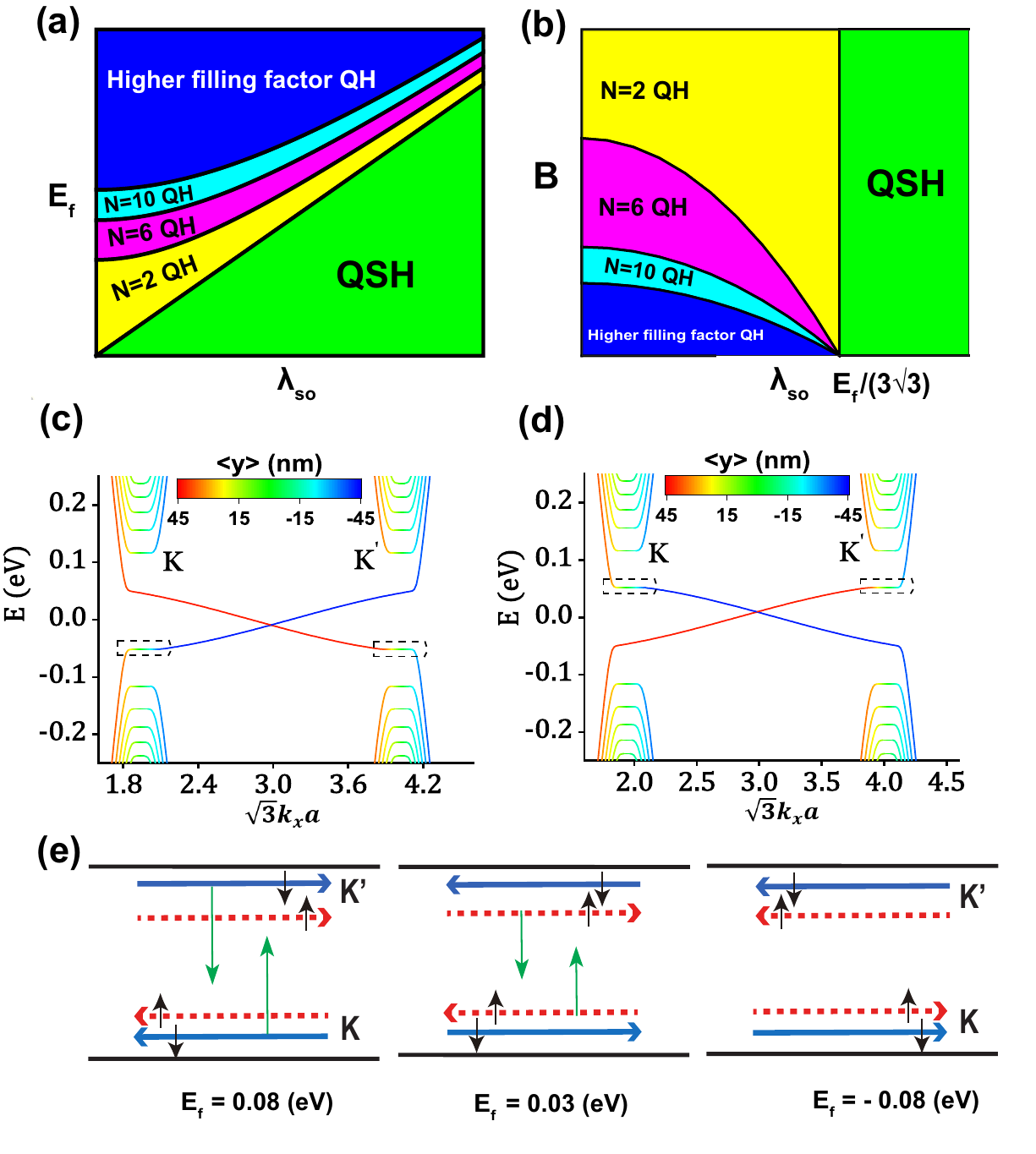}
	\centering 
	\caption{(a) The phase diagram for the QSH effect under a real magnetic field $B=5T$ as a function of $\lambda_{so}$ and Fermi level $E_{f}$. (b) The phase diagram for the QSH effect with a fixed $E_{f}$ as a function of $\lambda_{so}$ and $B$. (c, d) The energy bands for a zigzag graphene nanoribbon with $\lambda_{so}=0.01eV$, $B=10T$. The spin-up component for (c) and spin-down component for (d). The black dotted boxes circle the zeroth LL flat bands. (e) The schematic diagrams of edge states for different Fermi levels $E_{f}$. Red dotted arrow lines denote spin-up channels and blue solid arrow lines denote spin-down channels. The spin components are also labeled by black solid arrows. The green solid arrows denote the trend of edge states evolution when $E_f$ declines. $N_{y}=800$. }
	\label{FIG3}
\end{figure}

We further illustrate this effect from the perspective of phase diagrams. In Figs.~\ref{FIG3}(a,b), the phase diagram as a function of the Fermi level $E_{f}$ and SOC strength $\lambda_{so}$ with a fixed real magnetic field $B$ (a) as well as the phase diagram as a function of the real magnetic field $B$ and $\lambda_{so}$ with a fixed $E_{f}$ (b) are both shown. According to Table~\ref{Table1} and Ref \cite{Shevtsov2}, the phase boundaries should be described by
\begin{equation}
E_{f}=\sqrt{2e\hbar \vert B \vert n v_{f}^{2}+\Delta_{so}^{2}}
\end{equation}
with $(n=0,1,2,3,...)$. Due to valley and spin degeneracy in graphene \cite{Castro1}, the QH phases follow filling factors $N=2,6,10...$ as $E_{f}$ climbs with fixed $B$, $\lambda_{so}$ or as $B$ decreases with fixed $E_{f}$, $\lambda_{so}$. Here only low filling factor QH phases are considered (denoted by yellow, magenta, cyan patches). Other higher filling factor  QH phases are denoted by dark blue patches. As long as $E_{f}<\Delta_{so}$, the system can keep QSH phase (green patch) \cite{Shevtsov2, Beugeling}.

In Figs.~\ref{FIG3}(c,d), we plot the tight-binding energy bands for the QSH effect under a real magnetic field in a zigzag graphene nanoribbon for the spin-up component $(c)$ and spin-down component $(d)$. To exhibit the space evolution of the edge states, we compute the expectation values of the $\hat{y}$ position operator for all eigenstates, as shown by the color in Figs.~\ref{FIG3}(c,d). The edge states for three typical $E_{f}$ are schematically plotted in Fig.~\ref{FIG3}(e) (the red dotted arrow line denotes the spin-up channel while the blue solid arrow line denotes the spin-down channel). We emphatically pick out the zeroth LLs for $K/K'$ valley and spin-up/down component by black dotted boxes in Figs.~\ref{FIG3}(c,d). It is clear that the spin degeneracy of the zeroth LL is broken \cite{Martino,Shevtsov2}. The energy of the zeroth LL for the spin-up component is $-\Delta_{so}$ and of the zeroth LL for the spin-down component is $\Delta_{so}$, which are well consistent with the results in Table.~\ref{Table1}. Above the SOC energy gap $\Delta_{so}$, this is a typical QH phase with magnetic field-induced LLs \cite{Shevtsov2}. The edge states for two spin components from two valleys have the same chirality [e.g. $E=0.08 eV$ in Fig.~\ref{FIG3}(e)]. As $E_{f}$ goes down, the edge bands from the zeroth LL for the spin-down component will join the zeroth LL flat bands and therewith the edge states will emerge into bulk states [see black dotted boxes in Fig.~\ref{FIG3}(d)]. Conversely, edge states for the spin-up component stay at one edge since no LL flat bands will be crossed [see Fig.~\ref{FIG3}(c)]. We use green arrows to show the trend of edge states evolution [see Fig.~\ref{FIG3}(e), $E_{f}=0.08eV$]. Within the SOC energy gap [e.g. $E=0.03 eV$ in Fig.~\ref{FIG3}(e)], the QSH phase keeps \cite{Shevtsov2,Beugeling}. The spin-up and spin-down components contribute two edge states at both edges respectively with the opposite chirality. As $E_{f}$ continues to descend, edge bands from the zeroth LL for the spin-up component will join the zeroth LL flat bands [see black dotted boxes in Fig.~\ref{FIG3}(c)] and those edge states tend to emerge into bulk states [indicated by green arrows in Fig.~\ref{FIG3}(e), $E_{f}=0.03eV$], while edge states for the spin-down component remain still [see Fig.~\ref{FIG3}(d)]. Below the SOC energy gap [e.g. $E=-0.08 eV$ in Fig.~\ref{FIG3}(e)], a QH phase is recovered with a reversed chirality since carriers change from the electron type into the hole type. In short, as $E_{f}$ gradually decreases, the system transitions from the QH phase to the QSH phase and then to the QH phase again. During each transition, we find edge bands for only one spin component (i.e. so-called `unhappy' spin in Ref \cite{Shevtsov2}) cross the zeroth LL flat bands and their edge states leave from one edge into the bulk and further to the other edge, reversing the chirality.

Taking into account Zeeman effect, the spin-up energy band in Fig.~\ref{FIG3}(a) will be shifted up by $\Delta E_{z}=\frac{1}{2}g\mu_{B}B$ and the spin-down energy band in Fig.~\ref{FIG3}(b) will be shifted down by $\Delta E_{z}$. The QSH effect region within the zeroth LLs will be slightly narrowed but still persist within the energy range $[-\Delta_{so}+\Delta E_{z},\Delta_{so}-\Delta E_{z}]$. Since the Zeeman splitting is relatively small, the QSH effect remains in a large extent still. The edge states transition shown in Fig.~\ref{FIG3}(e) will not change.

\section{\label{sec4} The QSH effect under a pseudomagnetic field.}

In last section, from the QSH phase to the QH phase, time-reversal symmetry has been broken and only one spin component changes. For a strain-induced pseudomagnetic field, physical pictures should be different. Especially for edge states from the zeroth LL, they are a pair of spin-degenerate counterpropagating modes from two opposite valleys located at one edge as shown Fig.~\ref{FIG2}(a), reflecting the space inversion symmetry is broken. In this section, we analyze how a pseudomagnetic field affects the QSH effect.

\begin{figure}
	\includegraphics[width=1\columnwidth]{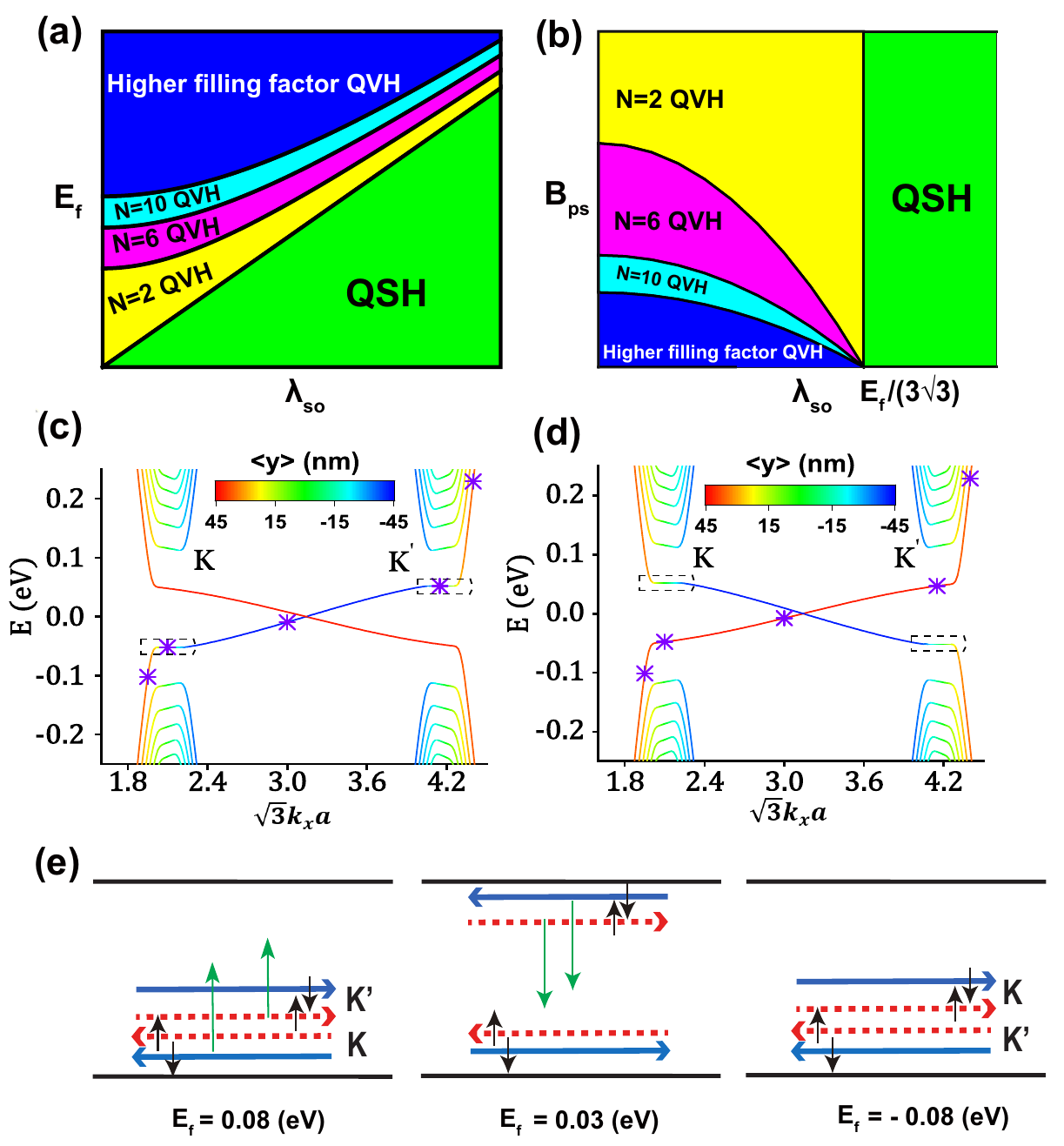}
	\centering
	\caption{(a) The phase diagram for the QSH effect under a pseudomagnetic field $B_{ps}=5T$ as a function of $\lambda_{so}$ and Fermi level $E_{f}$. (b) The phase diagram for the QSH effect with a fixed Fermi level $E_{f}$ as a function of $\lambda_{so}$ and pseudomagnetic field $B_{ps}$. (c, d) The energy bands for a zigzag graphene nanoribbon with $\lambda_{so}=0.01eV$, $B_{ps}=10T$. The spin-up component for (c) and spin-down component for (d). The black dotted boxes circle the zeroth LL flat bands. (e) The schematic diagrams of edge states for different Fermi levels $E_{f}$. Red dotted arrow lines denote spin-up channels and blue solid arrow lines denote spin-down channels. The spin components are also labeled by black solid arrows. The green solid arrows denote the trend of edge states evolution when $E_f$ declines. $N_{y}=800$. }
	\label{FIG4}
\end{figure}

\begin{figure}
	\includegraphics[width=1\columnwidth]{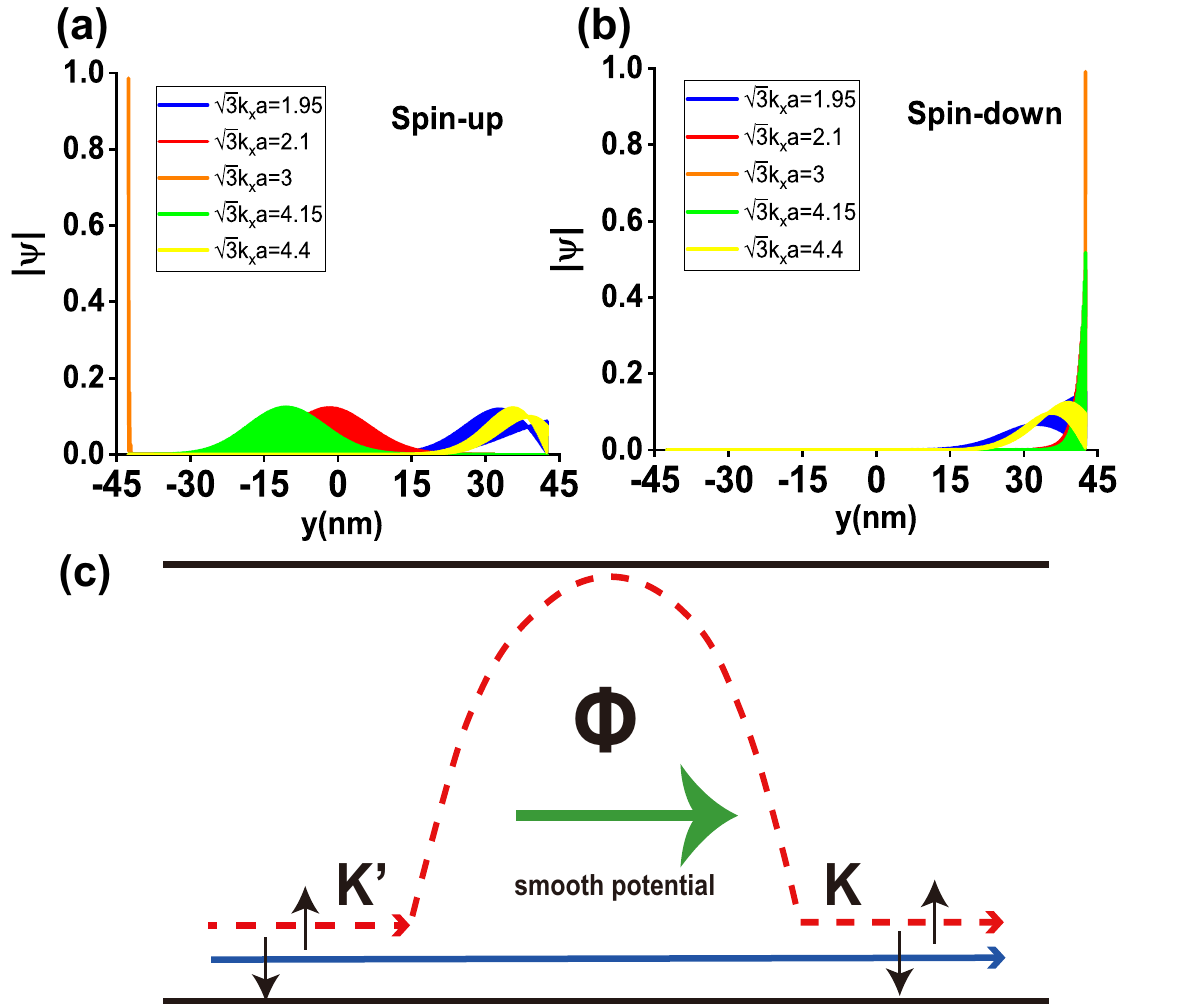}
	\centering 
	\caption{(a) and (b) respectively show the wavefunctions along one spin-up and one spin-down energy band for different $k_{x}$
corresponding to purple stars in Figs.~\ref{FIG4}(c) and (d).
(c) is the schematic diagram for the spin FET-like device based on the QSH effect under a pseudomagnetic field. The red dotted arrow line denotes the spin-up channel and blue solid arrow line denotes the spin-down channel.}
	\label{FIG5}
\end{figure}

In Figs.~\ref{FIG4}(a,b), we show the phase diagram as a function of Fermi level $E_{f}$ and SOC strength $\lambda_{so}$ with a fixed $B_{ps}$ (a) and the phase diagram as a function of $B_{ps}$ and $\lambda_{so}$ with a fixed $E_{f}$ (b). From a perspective of the phase distribution, the situations are basically the same as those under a real magnetic field in Figs.~\ref{FIG3}(a,b), except that QH phases are replaced by QVH phases.

For the tight-binding energy bands, some details become somehow different.
In Figs.~\ref{FIG4}(c,d),
we find the SOC energy gap still keeps like the real magnetic field case.
The QSH phase emerges in the SOC energy gap.
Outside the SOC energy gap, the space distributions of edge states
follow the QVH phase where two time reversal symmetry edge states stay
at the same edge [indicated by colors in Figs.~\ref{FIG4}(c,d)].
For each spin component, the zeroth LLs become no longer energy-degenerate
at two valleys [see black dotted boxes in Figs.~\ref{FIG4}(c,d)].
The energy of the zeroth LL for the spin-up (spin-down) component
at $K'$ valley ($K$ valley) is $\Delta_{so}$
while at $K$ valley ($K'$ valley) is $-\Delta_{so}$,
as Table~\ref{Table1} shows.
In other words, the zeroth LLs only appear in one branch of
decussate energy bands near the charge neutral point  [see the color branch in Figs.~\ref{FIG4}(c,d)]
rather than both. For the other branch, the states always stay at one edge since no LL flat bands exist [see the red branch in Figs.~\ref{FIG4}(c,d)].

These distinctions directly affect the evolution of edge states from the QVH phase to the QSH phase. When Fermi level $E_{f}$ is above the SOC energy gap, system is in the QVH phase and the counterpropagating edge states from the zeroth LLs stay at lower edge [e.g. $E=0.08eV$ in Fig.~\ref{FIG4}(e)].
As the $E_{f}$ declines, energy bands for the spin-up component at $K'$ valley and spin-down component at $K$ valley will join the zeroth LL flat bands [see black dotted boxes in Figs.~\ref{FIG4}(c,d)].
This suggests the time-reversal symmetry $\uparrow K'$ and $\downarrow K$ edge states will emerge into bulk states and then go to the upper edge of the nanoribbon [indicated by green arrows in left panel of Fig.~\ref{FIG4}(e)].
As a result, the QVH phase translates into the QSH phase.
Within the SOC energy gap, the helical edge states emerge.
The system is in the QSH phase [e.g. $E=0.03eV$ in Fig.~\ref{FIG4}(e)].
As $E_{f}$ continues to decrease, the edge bands at the upper edge
of the nanoribbon start to join the zeroth LL flat bands. These edge states evolve into bulk states again, and then go back to the lower edge.
Below the SOC energy gap, a QVH phase recovers and two pairs of time-reversal symmetry edge states from two opposite valleys stay at the lower edge again
[the right panel of Fig.~\ref{FIG4}(e)].

We can exhibit the evolution of the aforementioned edge states
through wavefunctions in Figs.~\ref{FIG5}(a,b).
These wavefunctions correspond to points along the rightward energy bands [purple stars in Figs.~\ref{FIG4}(c,d)]. For the spin-up component [Fig.~\ref{FIG5}(a)], the wavefunction clearly evolves from the lower edge of the nanoribbon
($\sqrt{3}k_{x}a=1.95$) to the bulk ($\sqrt{3}k_{x}a=2.1$),
then to the upper edge ($\sqrt{3}k_{x}a=3$),
next to the bulk ($\sqrt{3}k_{x}a=4.15$) again
and finally returns the lower edge ($\sqrt{3}k_{x}a=4.4$), matching Fig.~\ref{FIG4}(e) well.
While for the spin-down component [Fig.~\ref{FIG5}(b)], all wavefunctions always stay at the lower edge, due to this band no joining the zeroth LL flat band.
In addition, we also notice that the decay length of the QSH edge states is smaller than the LL edge states, because the decay length of the former depends on $\hbar v_{f}/\Delta_{so}$ while the latter depends on $l_{B}=\sqrt{\hbar/eB_{ps}}$ \cite{Shevtsov2}.

This picture where one spin channel runs from one edge to another edge could be used to conceive a spintronics device to adjust edge states spin direction. In Fig.~\ref{FIG5}(c), we apply a smooth potential gradient to design a p-n junction. Here only those rightward states are considered since smooth potential makes backscattering nearly unlikely to occur. In the p region and n region, the graphene is in the QVH phase. In the junction, the graphene is in the QSH phase. We also apply a weak real magnetic field on the junction (the coexistence of real magnetic field and pseudomagnetic field will not break our configuration and will be proved in the section VI). Based on previous discussions, one spin channel will surround the junction while the other will keep at one edge. See Fig.~\ref{FIG5}(c), if we inject a state with the spin polarized in the x direction:
\begin{equation}
	\psi_{i}=\frac{1}{\sqrt{2}}(\psi_{\uparrow}+\psi_{\downarrow}),
\end{equation}
where $\psi_{\uparrow/\downarrow}$ denotes the eigenstates with the spin-up/down polarized in the $z$ direction.

After traversing the junction, the spin-up channel $\psi_{\uparrow}$ surrounds the junction and accumulates a phase $\phi_{1}=\frac{1}{\hbar}\int_{l_{1}} (\vec{p}+e\vec{A})\cdot d\vec{r} $. The spin-down channel $\psi_{\downarrow}$ keeps at the lower edge and accumulates a phase $\phi_{2}=\frac{1}{\hbar}\int_{l_{2}} (\vec{p}+e\vec{A})\cdot d\vec{r}$.
Here $l_{1}$ and $l_{2}$ are the path for the spin-up and spin-down electrons to transverse the junction. So the outgoing wavefunction should be:
\begin{equation}
	\begin{split}
	\psi_{o}&=\frac{1}{\sqrt{2}}(\psi_{\uparrow}e^{i\phi_{1}}+\psi_{\downarrow}e^{i\phi_{2}})\\
	&=\frac{1}{\sqrt{2}}[\psi_{\uparrow}e^{i(\phi_{1}-\phi_{2})}+\psi_{\downarrow}]e^{i\phi_{2}} \\
	&=\frac{1}{\sqrt{2}}[\psi_{\uparrow}e^{i(\phi_{i}+2\pi \phi / \phi_{0})}+\psi_{\downarrow}]e^{i\phi_{2}}.
	\end{split}
	\label{Eq13}
\end{equation}
In the last row of Eq.~(\ref{Eq13}),
we divide the phase difference of $\phi_{1}-\phi_{2}$ into two parts:
$\phi_{i}$ is the phase accumulation at the zero real magnetic field
while $\phi$ is the flux from the real magnetic field since the two paths $l_{1}$ and $l_{2}$
just circle the junction.
So we can arbitrarily regulate the relative phase difference between $\psi_{\uparrow}$ and $\psi_{\downarrow}$ by adjusting the magnetic flux $\phi$.
In other words, just by varying the magnetic flux,
we can continuously rotate the spin polarization direction in the x-y plane of the outgoing state $\psi_{o}$.
Our device actually has a similar function as a spin field effect transistor (spin-FET) or Datta-Das transistor which is an electronic analog of the electro-optic modulator \cite{Datta}.
The difference is that we use a QVH-QSH-QVH junction to implement the spin-related Aharonov-Bohm effect \cite{Aharonov} to control states spin precession instead of the Rashba SOC.

\section{\label{sec6} The QSH effect under the coexistence of a real magnetic and a pseudomagnetic field.}

\begin{figure*}
	\includegraphics[width=0.8\textwidth]{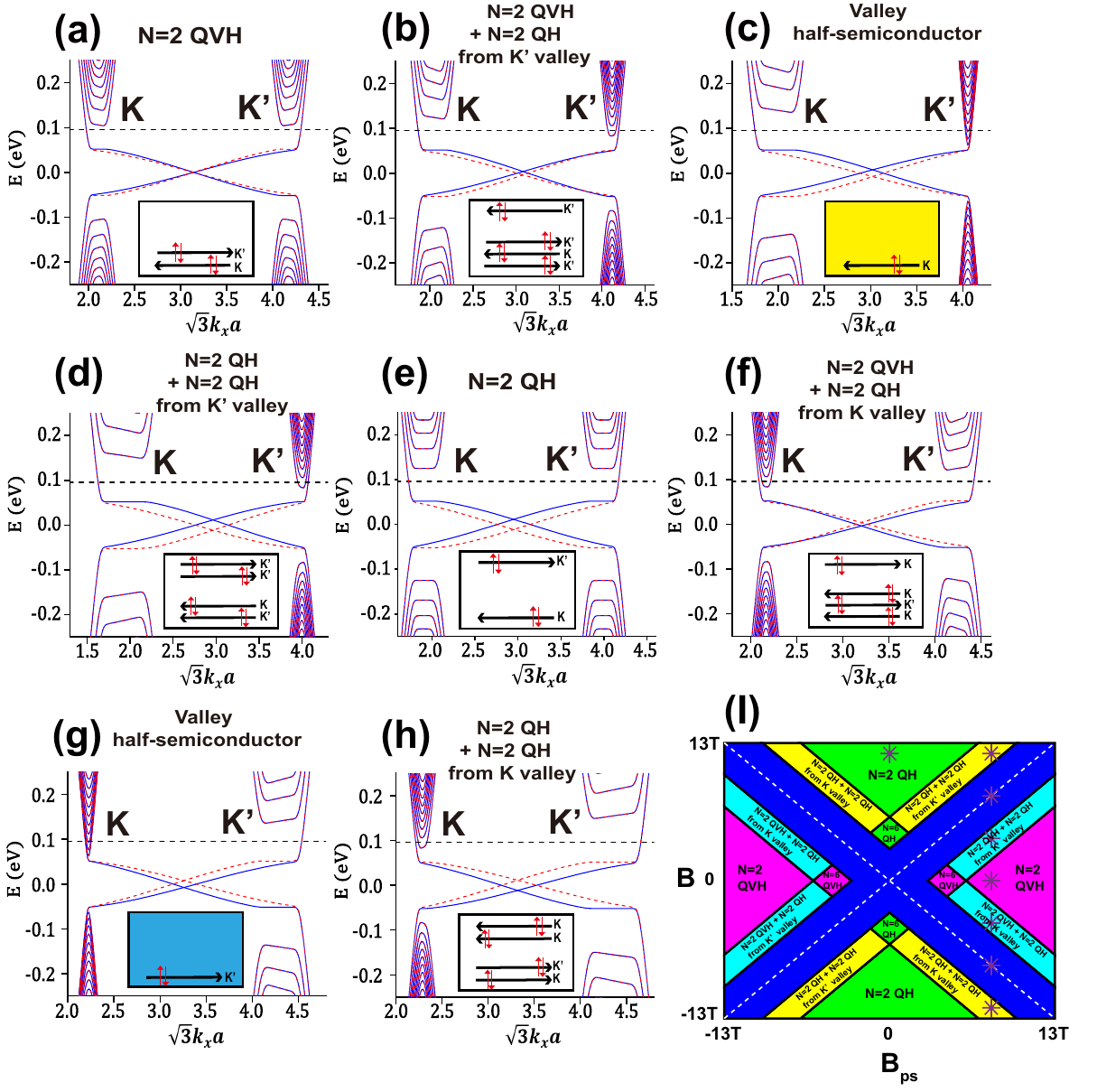}
	\centering 
	\caption{(a-h) The energy bands for a zigzag graphene nanoribbon with $\lambda_{so}=0.01eV$ where $B$ and $B_{ps}$ coexist. The red dotted lines denote the spin-up component and the blue solid lines denote the spin-down component. The black dotted lines denote the Fermi level $E_{f}=\sqrt{2e \hbar v_{f}^{2}*6+\Delta_{so}^{2}}\approx 0.0958eV$. The insets in (a-h) schematically shows the edge states distributions by black solid arrows at Fermi level $E_{f}$. The red solid arrows on edge states also label spin-up and spin-down components. The yellow color in (c) denotes bulk states from $K'$ valley and the blue color in (g) denotes the bulk states from $K$ valley.
	The parameters (a) $B_{ps}=8T,B=0T$, (b) $B_{ps}=8T,B=4T$, (c) $B_{ps}=8T,B=8T$, (d) $B_{ps}=8T,B=12T$, (e) $B_{ps}=0T,B=12T$,
	(f) $B_{ps}=8T,B=-4T$, (g) $B_{ps}=8T,B=-8T$, and (h) $B_{ps}=8T,B=-12T$. (I) The phase diagram as a function of $B$ and $B_{ps}$ with the same $E_{f}>\Delta_{so}$. The purple stars in (I) correspond to (a-h). The width of the graphene nanoribbon $N_{y}=800$.  }
	\label{FIG6}
\end{figure*}

\begin{figure*}
	\includegraphics[width=0.8\textwidth]{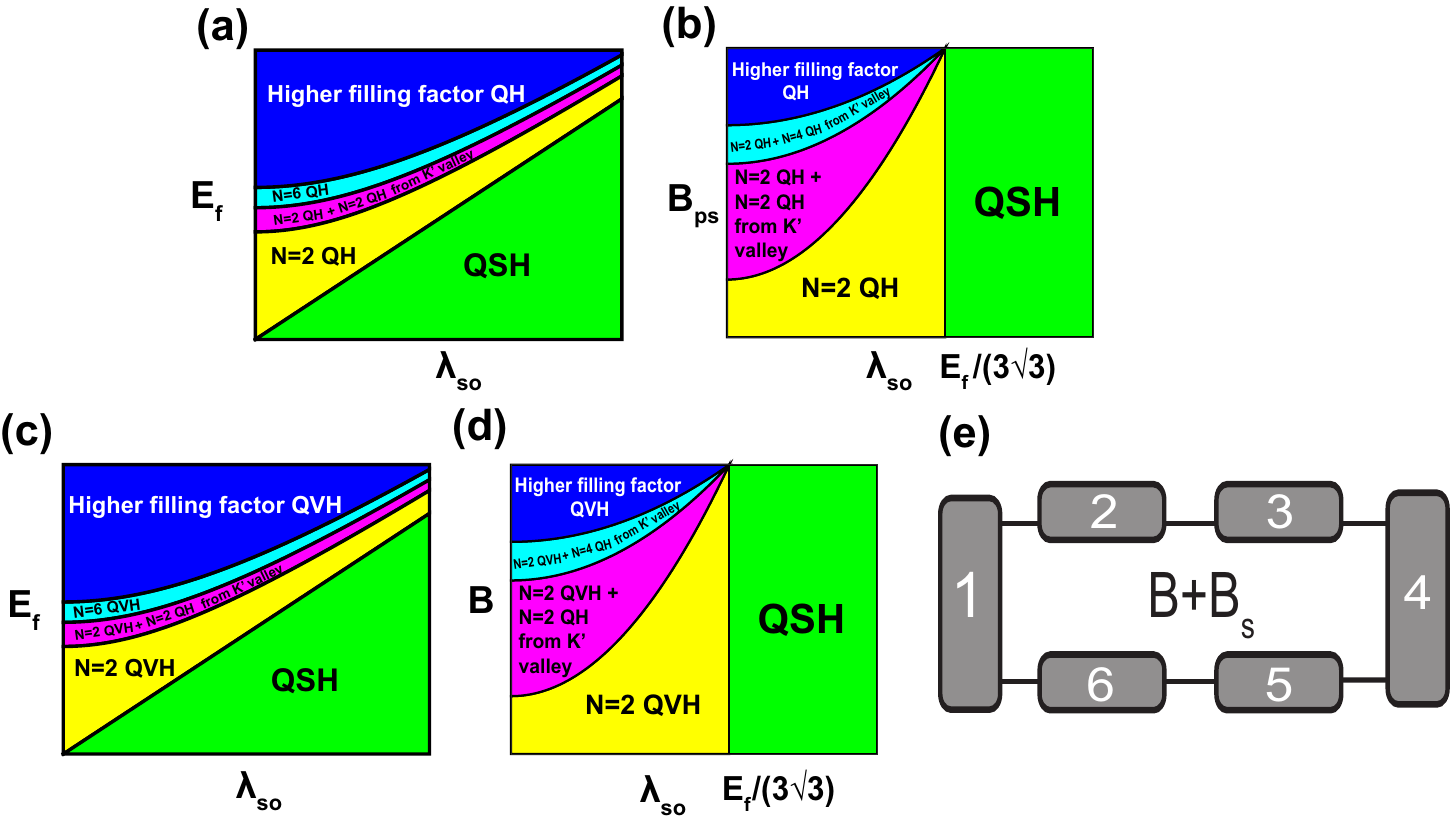}
	\centering 
	\caption{(a-b) $B$-dominated phase diagrams as a function of $\lambda_{so}$ and $E_{f}$ with fixed $B=8T,B_{ps}=2T$ (a) and as a function of $\lambda_{so}$ and $B_{ps}$ with fixed $E_{f},B=10T$ (b). (c-d) $B_{ps}$-dominated phase diagrams as a function of $\lambda_{so}$ and $E_{f}$ with fixed $B=2T,B_{ps}=8T$ (c) and as a function of $\lambda_{so}$ and $B$ with fixed $E_{f},B_{ps}=10T$ (d). (e) The schematic diagram for a six-terminal device. }
	\label{FIG7}
\end{figure*}

\begin{figure}
	\includegraphics[width=1\columnwidth]{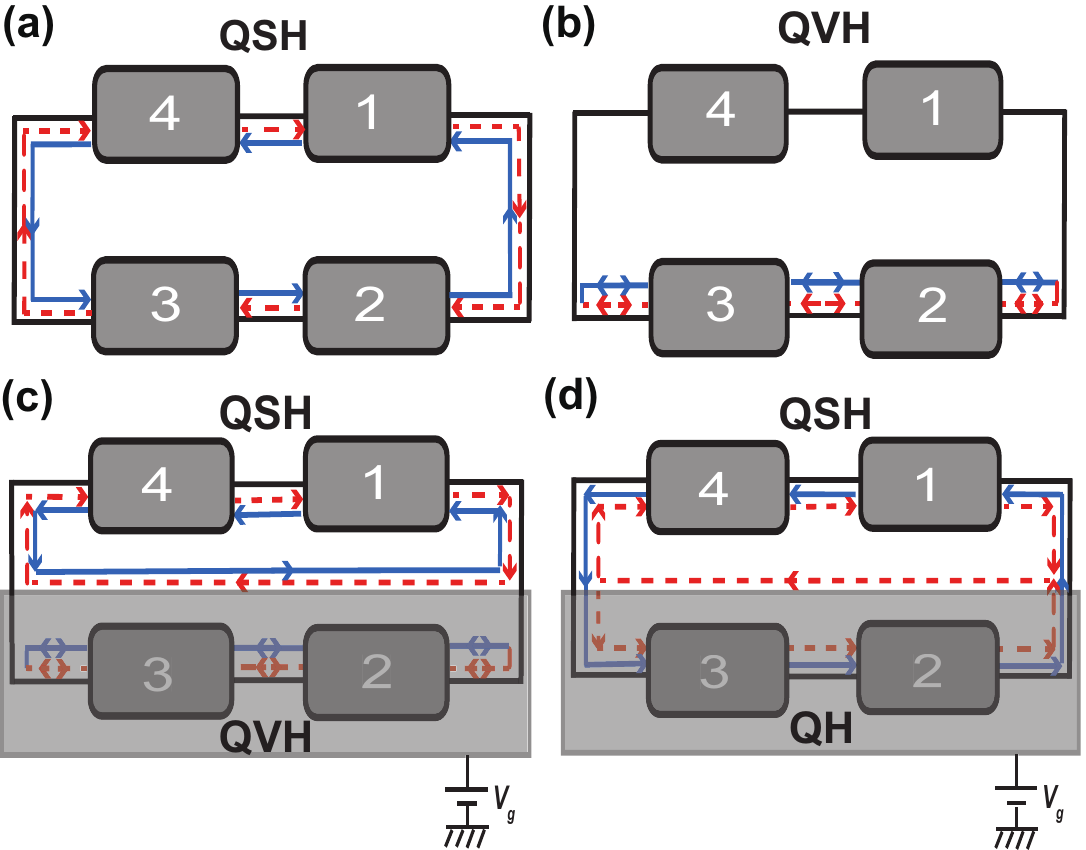}
	\centering 
	\caption{The schematic diagrams of edge states transport for various configurations in a four-terminal device. (a) The system is in the QSH phase. (b) The system is in the QVH phase. (c) The system constitutes a junction of the QSH phase and QVH phase by a top gate on the lower half plane. (d) The system constitutes a junction of the QSH phase and QH phase by a top gate on the lower half plane and with a real magnetic field. The red dotted lines denote spin-up channels and the blue solid lines denote spin-down channels.}
	\label{FIG8}
\end{figure}
In previous sections, we have clarified the relationships between the real magnetic field and the pseudomagnetic field, the real magnetic field and the QSH effect, as well as the pseudomagnetic field and the QSH effect. Now we turn to figure out how these three phases compete and coexist. We still pay attention to phase diagrams, energy bands and edge states. Finally, we design a spintronics multiple-way switch based on the relationship of these three effects.

Referring to Table~\ref{Table1}, we naturally generalize the energy of LLs from $K$ valley and $K'$ valley taking into account the SOC $\lambda_{so}$:
 \begin{equation}
	\begin{split}
	E_{\pm n}^{K}&=\pm \sqrt{2 e \hbar \vert B+B_{ps} \vert  v_{f}^{2}n+\Delta_{so}^{2}} \\
	E_{\pm n}^{K'}&=\pm \sqrt{2 e \hbar \vert B-B_{ps} \vert  v_{f}^{2}n+\Delta_{so}^{2}},
	\end{split}
	\label{Eq14}
 \end{equation}
with $n=1,2,...$. The energy of the zeroth LLs for each valley is still $\pm \Delta_{so}$ with the sign depending on the sign of the total magnetic field.   In Figs.~\ref{FIG6}(a-h), we show the energy bands of a zigzag graphene nanoribbon with $\lambda_{so}$, $B$ and $B_{ps}$. We also schematically plot edge states distributions in the insets at Fermi level $E_{f}=\sqrt{2e \hbar v_{f}^{2}*6+\Delta_{so}^{2}}\approx 0.0958eV$ denoted by black dotted lines in Figs.~\ref{FIG6}(a-h). Although these results, especially for edge states distributions, are parallel to Fig.~\ref{FIG2}, some points or discrepancies should be noticed:
(1) On the one hand, a SOC energy gap $\Delta_{so}$ opened by $\lambda_{so}$ is hardly affected by the real magnetic field $B$
and the pseudomagnetic field $B_{ps}$.
The phase of the system depends on where the Fermi level $E_{f}$ stays.
When $E_f$ is within the SOC energy gap, the system is in the QSH phase with a pair of helical edge states, which is independent of $B$ and $B_{ps}$.
(2) On the other hand, when $E_f$ is outside the SOC energy gap,
the system is in the QH or the QVH phase depending on
the real magnetic field $B$ or the pseudomagnetic field $B_{ps}$.
As $B$ or $B_{ps}$ changes,
the system evolves between QH and QVH phases [see Figs.~\ref{FIG6}(a-h)].
However, from the QH phase to the QVH phase and vice versa,
the energy band does not experience a global energy gap close
and open. This is quite different from the case of $\lambda_{so}=0$ [see Fig.~\ref{FIG2}].
In other words, $B$ and $B_{ps}$ only affect
the energy bands outside the SOC energy gap.
(3) At the special points $\vert B \vert = \vert B_{ps} \vert$,
the total magnetic field at one valley disappears.
Due to the existence of SOC energy gap,
this peculiar energy band should be regarded as a valley half-semiconductor rather than the valley halfmetal
[see Figs.~\ref{FIG6}(c,g) and Figs.~\ref{FIG2}(c,g)].
In comparison, such a valley half-semiconductor would have stronger
application value than a valley halfmetal. Because the valley half-semiconductor has an energy gap, the carrier density can be well controlled.
We can adjust the Fermi level $E_{f}$ to achieve
a complete valley-polarized transport ($E_{f}$ above the SOC energy gap)
and also a spin-related transport ($E_{f}$ within the SOC energy gap).
(4) The energy bands and edge states above the SOC energy gap can approximately maintain the spin degeneracy, but within the SOC gap the spin degeneracy has been strongly destroyed,
since time reversal symmetry and space inversion symmetry are both broken.

The energy bands in Fig.~\ref{FIG6} clearly show that
when $B$, $B_{ps}$ and $\lambda_{so}$ coexist,
$B$ and $B_{ps}$ just compete with each other
outside the SOC energy gap. As long as $\lambda_{so}$ persists,
the QSH phase within the SOC energy gap will not be disturbed.
Therefore, the phase diagram as a function of $B$ and $B_{ps}$ in
Fig.~\ref{FIG6}(i) with Fermi energy fixed at $E_{f}>\Delta_{so}$
should be identical to Fig.~\ref{FIG1}.

Figures.~\ref{FIG7}(a-d) show the phase diagrams as a function
of $E_{f}$ and $\lambda_{so}$, or $B$ ($B_{ps}$) and $\lambda_{so}$.
Here we still divide the situations into two regions \cite{Roy2}:
$B$-dominated [Figs.~\ref{FIG7}(a,b)] and
$B_{ps}$-dominated [Figs.~\ref{FIG7}(c,d)].
The phases boundaries are coordinated with Eq.~(\ref{Eq14}).
In Figs.~\ref{FIG7}(a,c), the phase diagram is as a function of $E_{f}$ and $\lambda_{so}$, with a fixed $B$ and $B_{ps}$.
$E_{f}=\Delta_{so}$ separates the QSH phase and the QH phase $(B > B_{ps})$
or the QVH phase $(B < B_{ps})$.
Unlike Fig.~\ref{FIG3}(a) and Fig.~\ref{FIG4}(a),
an additional valley-imbalanced phase,
`$N$=2 QH (QVH) + $N$=2 QH from $K'$ valley'
[see the pink patches in Figs.~\ref{FIG7}(a,c)], emerges
when $E_{f}$ is between $E_{1}^{K}$ and $E_{1}^{K'}$,
since one extra spin-degenerate LL from $K'$ valley is crossed.
Similar situations appear in phase diagrams as a function of $B_{ps}$ ($B$)
and $\lambda_{so}$ as well [see Figs.~\ref{FIG7}(b,d)].
If the $\lambda_{so}$ is constant and $B_{ps}$ ($B$) climbs,
according to the Equation~(\ref{Eq14}), more and more LLs from $K'$
valley will descend and be crossed, echoing valley-imbalanced phases,
e.g. the `$N$=2 QH (QVH) + $N$=2 (4) QH from $K'$ valley' phase denoted by magenta and cyan in Figs.~\ref{FIG7}(b,d).

Below we also illustrate the edge states distributions
of these phases in Figs.~\ref{FIG6} and \ref{FIG7}.
For the QSH phase, a pair of helical edge states with opposite spin channels
propagate in opposite directions \cite{Hasan, Qi}.
The $N$=2 QH, $N$=2 QVH, `$N$=2 QH + $N$=2 QH from $K'/K$ valley'
and `$N$=2 QVH + $N$=2 QH from $K'/K$ valley' phase
follow the edge state distributions in Figs.~\ref{FIG6}(e), (a),
(d)/(h), and (b)/(f).
For the $N$=6 QH phase, three chiral spin-degenerate edge states
appearing at both edges analogy to Fig.~\ref{FIG6}(e).
For the $N$=6 QVH phase, two pairs of counterpropagating
spin-degenerate edge states appear at the lower edge
while one pair of counterpropagating spin-degenerate
edge states appear at the upper edge.
For the `$N$=2 QH + $N$=4 QH from $K'/K$ valley' phase,
three chiral spin-degenerate edge states propagate
in opposite directions at opposite edges, analogy to Fig.~\ref{FIG6}(d).
For the `$N$=2 QVH + $N$=4 QH from $K'$ valley' phase,
there are two leftward spin-degenerate edge states at the upper edge,
and one leftward spin-degenerate edge states
as well as three rightward spin-degenerate edge states at the lower edge.

From the perspective of the transport,
we can characterize these aforementioned phases
in terms of longitudinal resistance $R_{xx}$ and Hall resistance $R_{H}$,
by considering a six-terminal device as shown in Fig.~\ref{FIG7}(e).
In the linear response regime, the current flowing from the $p$ terminal is calculated by B$\ddot{u}$ttiker formula:\cite{Buttiker}
\begin{equation}
	I_{p}=\frac{e^{2}}{h}\sum_{q}T_{pq}(V_{p}-V_{q}),
	\label{Eq15}
\end{equation}
where $V_{p}$ is the voltage in the terminal $p$
and $T_{pq}$ is the transmission coefficients from terminal $q$ to terminal $p$.
In theoretical calculations, we set that the transmission coefficients $T_{pq}$
are equal to the number of the edge states from terminal $q$ to terminal $p$
by neglecting disorder-induced backscattering.
A voltage $V$ is applied on the terminal 1. The terminal 4 is grounded.
Other terminals are voltage terminals with zero currents.
Along with these conditions, the currents and voltages for each terminal can be obtained by directly solving Equation~(\ref{Eq15}).
We note that because space inversion symmetry has been broken by $B_{ps}$,
$R_{xx}$ and $R_{H}$ are better to define relying on which edge.
For the upper edge, $R_{xx}^{u}=\frac{V_{2}-V_{3}}{I_{1}}$.
For the lower edge, $R_{xx}^{d}=\frac{V_{6}-V_{5}}{I_{1}}$.
The Hall resistances are $R_{H}^{L}=\frac{V_{2}-V_{6} }{I_{1}}$
and $R_{H}^{R}=\frac{V_{3}-V_{5} }{I_{1}}$.

Based on explicit edge modes distributions in Fig.~\ref{FIG6},
we can straightly present the calculation results.
In the QSH phase,  $R_{xx}^{u}=R_{xx}^{d}=\frac{h}{2e^{2}}$ and $R_{H}^{L}=R_{H}^{R}=0$ \cite{Hasan,Qi,Chen}. For the $N$=2 QH phase in Fig.~\ref{FIG6}(e), the $R_{xx}^{u}=R_{xx}^{d}=0$ and $R_{H}^{L}=R_{H}^{R}=\frac{h}{2e^{2}}$. For the $N$=2 QVH phase in Fig.~\ref{FIG6}(a), $R_{xx}^{d}=\frac{h}{2e^{2}}$ \cite{Wu}.
Because edge states of the $N$=2 QVH phase
only appear at the lower edge, terminals 2 and 3 are disconnected with the edge states.
The voltages $V_2$ and $V_3$ are strongly affected by disorders and hard to be determined,
so $R_{H}^{L}$, $R_{H}^{R}$ and $R_{xx}^{u}$ should be uncertain and depend on
disorder configurations.
For the `$N$=2 QH + $N$=2 QH from $K'/K$ valley' phase in Figs.~\ref{FIG6}(d,h), $R_{xx}^{u}=R_{xx}^{d}=0$. $R_{H}^{L}=R_{H}^{R}=\frac{h}{4e^{2}}$ for $(d)$ and $-\frac{h}{4e^{2}}$ for $(h)$.
For the `$N$=2 QVH + $N$=2 QH from $K'$ valley' phase [Fig.~\ref{FIG6}(b)],
$R_{xx}^{u}=0$, $R_{xx}^{d}=\frac{h}{8e^{2}}$, $R_{H}^{L}=-\frac{3h}{8e^{2}}$ and $R_{H}^{R}=-\frac{h}{4e^{2}}$.
Similarly, for the `$N$=2 QVH + $N$=2 QH from $K$ valley' phase [Fig.~\ref{FIG6}(f)], $R_{xx}^{u}=0$, $R_{xx}^{d}=\frac{h}{8e^{2}}$, $R_{H}^{L}=\frac{h}{4e^{2}}$ and $R_{H}^{R}=\frac{3h}{8e^{2}}$.

In Ref \cite{Shevtsov2}, the QSH phase with a real magnetic field allows
for a direct junction between the QSH phase and the QH phase via an additional electrostatic gate.
The different chirality for one spin channel will compel it to propagate along the interface.
This system thus provides a very efficient spin-polarized charge-current switching mechanism.
Compared with the spin flip from the QSH phase to the QH phase, the biggest feature from the QSH phase to the QVH phase is a ``side flip" as shown in Fig.~\ref{FIG4}(e).
Enlighten by this, we can use a junction between the QSH phase and the QVH phase to achieve a spintronics multiple-way switch.
Considering a four-terminal device, the edge states of the QSH phase ($\lambda_{so} \neq 0, B_{ps}=0$) in Fig.~\ref{FIG8}(a) and the QVH phase ($\lambda_{so}=0, B_{ps}\neq 0$) in Fig.~\ref{FIG8}(b) are shown respectively. A pair of helical edge states with opposite spins propagating in opposite direction appear at both edges in the QSH phase. The counterpropagating spin-degenerate edge states only appear at the lower edge in the QVH phase.
We apply a voltage on terminal 1 ($V_{1}=V$) and ground all the other terminals $(V_{2}=V_{3}=V_{4}=0)$.
In the QSH phase [Fig.~\ref{FIG8}(a)], the spin-down current (blue solid arrow lines) will flow from the terminal 1 to the terminal 4 and the spin-up current (red dotted arrow lines) will flow from the terminal 1 to the terminal 2. In the QVH phase [Fig.~\ref{FIG8}(b)], the upper edge is disconnected and no current flows.
Next we combine both effects $(\lambda_{so} \neq 0, B_{ps} \neq 0)$ in Fig.~\ref{FIG8}(c).
If the Fermi level $E_{f}$ is within the SOC energy gap, the system is still in the QSH phase like Fig.~\ref{FIG8}(a). When a local top gate is applied on the lower half plane and drives the lower half plane (covered by gray plane) into the QVH phase by adjusting $E_{f}$ above the SOC energy gap, a direct QSH-QVH junction forms [Fig.~\ref{FIG8}(c)]. Since the QVH phase only has channels at the lower edge, the spin-up edge flow of the QSH phase cannot flow into the lower half plane and must flow into the terminal 4 through the interface. Hence, by gating the lower half plane into the QSH phase or the QVH phase, we can control the spin-up current from the terminal 1 to flow into the terminal 2 or terminal 4. Furthermore, if a magnetic field is applied on this QSH-QVH junction $(\lambda_{so} \neq 0, \vert B \vert > B_{ps} \neq 0, B < 0)$ to magnetize the QVH phase in the lower half plane into the QH phase (the QSH phase will not be interrupted), the spin-up chiral state will emerge along the interface \cite{Shevtsov2} and flow into the terminal 3 and 4 [Fig.~\ref{FIG8}(d)]. In short, using a local gate and magnetic field on the graphene with SOC and pseudomagnetic field, we can realize a direct junction of QSH-QVH or QSH-QH. This implements a multiple-way spintronics switch to control one spin-polarized current flowing from the terminal 1 into terminals 2,3 and 4.

\section{\label{sec7} Conclusion.}
In summary, we investigate the phase diagrams and edge states transitions
when the QH effect, the QSH effect and the QVH effect coexist in graphene.
We find the QSH effect can still keep within the SOC energy gap
under a relatively large real magnetic field and pseudomagnetic field.
Outside the SOC energy gap, the real magnetic field and pseudomagnetic field
will compete and affect the LLs at each valley.
Moreover, the zeroth LLs of different valleys and spins
exhibit different distributions for the real magnetic field
and pseudomagnetic field, leading to distinctive edge states transitions
from the QSH phase to the QH phase and to the QVH phase.
Specially in the case of the pseudomagnetic field,
the spatially asymmetric edge state structure
can be used to construct a spin FET-like device to tune
the spin polarization of edge state transport
and realize a multiple-way spintronics switch.

\section*{Acknowledgments.}
This work was financially supported by NSF-China (Grant No. 11921005), National Key R and D program of China (Grant No. 2017YFA0303301), and the Strategic Priority Research Program of Chinese Academy of Sciences (Grant No. XDB28000000).
\\
\\

\begin{figure}
	\includegraphics[width=1\columnwidth]{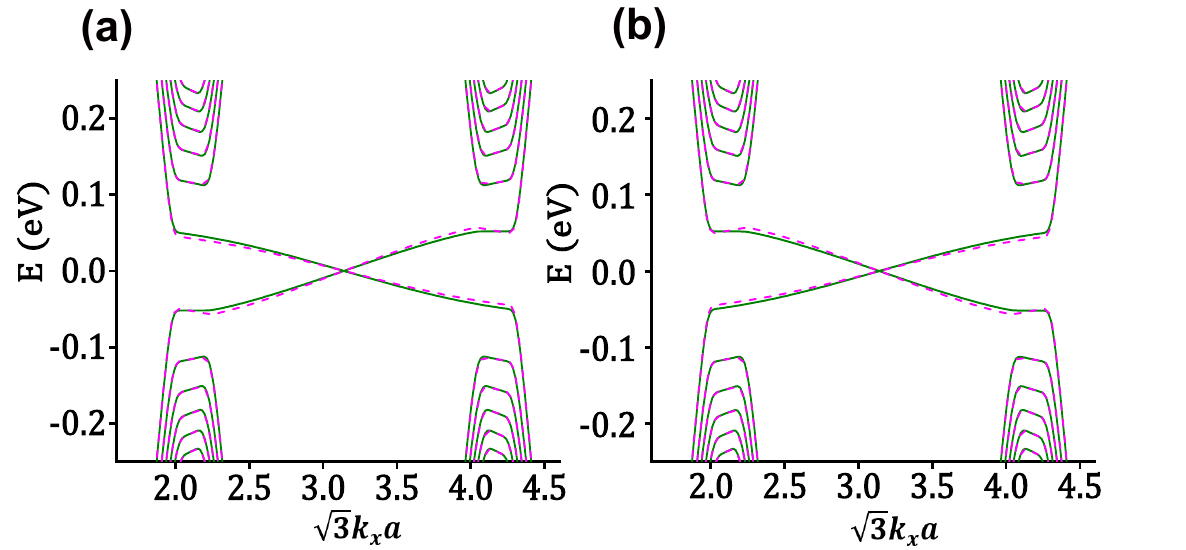}
	\centering 
	\caption{The comparisons of the energy bands under $B_{ps}=10T$, $\lambda_{so}=0.01eV$ between the case without SOC modulation (olive green solid lines) and the case with SOC modulation (magenta dotted lines). (a) The spin-up component. (b) The spin-down component. Here $\gamma=3,\alpha=1.8$, $N_{y}=800$.}
	\label{FIG9}
\end{figure}

\appendix
\section{The effect of strain on the strength of intrinsic SOC}

In this appendix, we consider how the strain field modifies the strength for intrinsic SOC $\lambda_{so}$ and inspect whether it could affect our results in the main text.

In Eq.~(\ref{Eq2}), we directly modify the nearest-neighbor hopping parameters $t_{ij}$ by an exponential relation. For SOC, the situation will be subtle. The intrinsic SOC in graphene originates from the intra-atomic SOC which allows for the transition between the $\pi$ band and $\sigma$ band near the Dirac points \cite{Hernando}. Therefore, to include the SOC modifications in the $\pi$ band effective Hamiltonian like Equation~(\ref{Eq2}), the multiple transfer process from $\pi$ to $\sigma$ bonds must be considered. From the previous literature, the lowest-order contribution is the next-nearest-neighbor hopping, i.e. passing from $\pi$ to $\pi$ through $\sigma$ bonds, and the $\lambda_{so}$ can be derived as: \cite{Min,Yao}
\begin{equation}
	\lambda_{so} \approx \frac{\Delta \xi_{0}^{2}}{18 V_{sp\sigma}^{2}},
	\label{EqA1}
\end{equation}
where $\Delta$ is the energy difference between the $2s$ and $2p$ orbitals, $\xi_{0}$ is the intra-atomic SOC constant, $V_{sp\sigma}$ is the Slater-Koster (SK) parameter formed by $2s$ and $2p$ orbitals.

The strain field will change the distance of bonds between lattices and thus correct the $V_{sp\sigma}$. Usually, the variation of SK parameters with the strain can be modeled as an exponential function: \cite{Pereira2}
\begin{equation}
	V_{sp\sigma}=V_{sp\sigma}^{0}e^{-\alpha(d/a-1)}
	\label{EqA2}
\end{equation}
where $V_{sp\sigma}^{0}$ is the unstrained SK parameter, $\alpha$ is
a parameter to measure the effect of the strain on the SK parameters and $d$ is the distance of the bond formed by $2s$ and $2p$ orbitals. We can regard the strain as a small perturbation so the Equation~(\ref{EqA1}) is still valid. Putting Equation~(\ref{EqA2}) into Equation~(\ref{EqA1}), we can derive a relation:
\begin{equation}
	\lambda_{so,ij} \approx \lambda_{so}^{0}e^{\alpha(d_{il}/a+d_{lj}/a-2)}.
	\label{EqA3}
\end{equation}
 Here $\lambda_{so,ij}$ is the SOC strength depending on the next-nearest-neighbor lattice indices $i$ and $j$. $l$ denotes the middle lattice site when passing from $j$ to $i$. $\lambda_{so}^{0}$ is the unstrained SOC strength. Equation~(\ref{EqA3}) is very similar to the Equation~(\ref{Eq2}) except that the minus sign in the exponential changes into the plus sign. $\alpha$ takes the role of $\gamma$ in Equation~(\ref{Eq2}) with a value around 2 \cite{Rezaei}.

Now we investigate if $\lambda_{so}$ is modified as $\lambda_{so,ij}$ in Equation~(\ref{EqA3}), what the SOC term $H_{so}$ (the second term in Equation~(\ref{Eq1})) could bring about. For each lattice point, we denote its six next-nearest-neighbor SOC strength as $\lambda_{so,j} (j=1,2,3,..6)$.  Using the Fourier transformation, $H_{so}$ can be transformed into the Brillouin zone under the basis $(\phi_{A \uparrow},\phi_{A \downarrow},\phi_{B \uparrow},\phi_{B \downarrow} )$ :
\begin{equation}
	\begin{split}
	&H_{so}(k)= \begin{pmatrix}
    H^{A}_{so}(k)  & 0 \\
	0 & H^{B}_{so}(k)
	\end{pmatrix} \\
	&= \begin{pmatrix}
		\sum_{j, A}i\lambda_{so,j}\upsilon_{j}s_{z}e^{i\vec{k}\cdot \vec{D}_{j}}  & 0 \\
	0 & \sum_{j, B}i\lambda_{so,j}\upsilon_{j}s_{z}e^{i\vec{k}\cdot \vec{D}_{j}}
	\end{pmatrix}.
    \end{split}
	\label{EqA4}
\end{equation}
Here $j$ denotes the next-nearest-neighbor bond and $\vec{D}_{j}$ is the bond vector of the next-nearest-neighbor bond $j$ \cite{refa4}. Using the linear expansion $\lambda_{so,j} \approx \lambda_{so}^{0}(1+\delta\lambda_{so,j})$ and the relation $\vec{D}_{j}=( I+\bar{\epsilon}) \vec{D}_{j}$, we can expand $H^{A}_{so}$ near the Dirac point $\vec{k}=\vec{K}+\vec{q}$:

\begin{equation}
\begin{split}
	H^{A}_{so}&=\sum_{j, A} i\lambda_{so,j}\upsilon_{j}s_{z}e^{i(\vec{K}+\vec{q}) \cdot \vec{D}_{j}}\\
	&\approx \sum_{j, A} i(\lambda_{so}^{0}+\lambda_{so}^{0}\delta\lambda_{so,j})\upsilon_{j}s_{z}e^{i(\vec{K}+\vec{q}) \cdot (\vec{D}_{j}+\bar{\epsilon} \vec{D}_{j})}\\
	&\approx \sum_{j, A} i(\lambda_{so}^{0}+\lambda_{so}^{0}\delta\lambda_{so,j})\upsilon_{j}s_{z}e^{i\vec{K} \cdot \vec{D}_{j}} (1+\vec{K} \cdot \bar{\epsilon} \vec{D}_{j}+\vec{q} \cdot \vec{D}_{j})\\
	&\approx \sum_{j, A} i\lambda_{so}^{0}\upsilon_{j}s_{z}e^{i\vec{K} \cdot \vec{D}_{j}}+\sum_{j, A} i\lambda_{so}^{0}\delta\lambda_{so,j}\upsilon_{j}s_{z}e^{i\vec{K} \cdot \vec{D}_{j}}\\
	&+\sum_{j, A} i\lambda_{so}^{0}\vec{K} \cdot \bar{\epsilon} \vec{D}_{j}\upsilon_{j}s_{z}e^{i\vec{K} \cdot \vec{D}_{j}}.
\end{split}
\label{EqA5}
\end{equation}

In the last equality of Equation~(\ref{EqA5}), we ignore the contribution from $\vec{q}$ which is a small amount around the Dirac point. The first term of the last equality is the zeroth contribution of the SOC which is just a mass term $\Delta_{so}s_{z}$ as Equation~(\ref{Eq9}) indicates. The second term is the first correction of SOC and should be tiny since $\delta\lambda_{so,j}$ is generally smaller than $10\%$. The third term should be also neglected as it is small too and irrelevant to SOC strength modulation. Therefore, the modification from strain on the SOC strength should not affect our results, in view that the main contributions come from the SOC zeroth contribution. This is different from the case for the nearest-neighbor hopping modulations where the zeroth contributions disappear at the Dirac points. As for $H_{so}^{B}$ and $K'$ valley, situations are alike. Furthermore, in Figs.~\ref{FIG9}(a,b), we plot the energy bands $(B_{s}=10T, \lambda_{so}=0.01eV)$ with SOC corrections (magenta dotted lines) and without SOC corrections (olive green solid lines). It is clear that results are almost identical.

\begin{figure}
	\includegraphics[width=1\columnwidth]{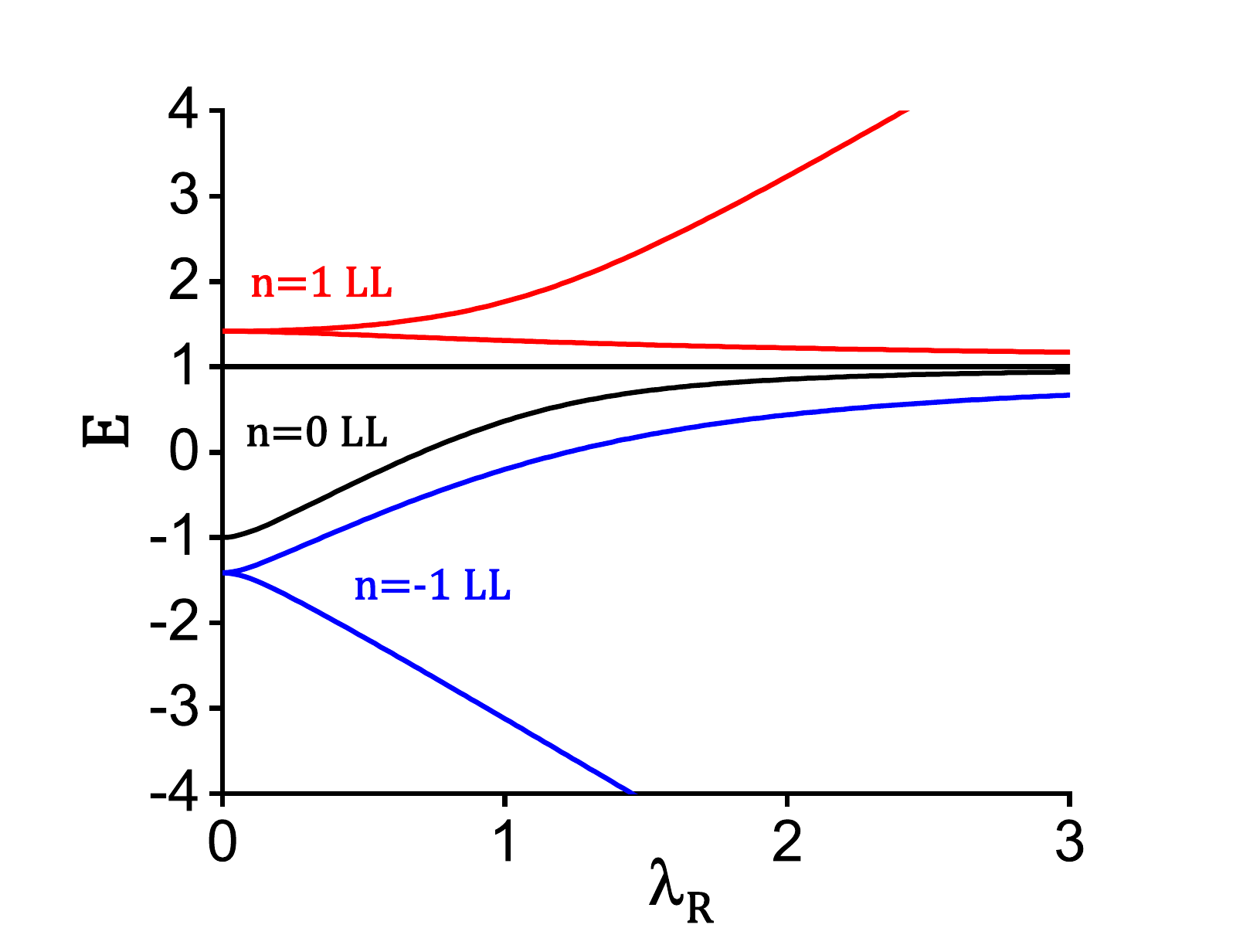}
	\centering 
	\caption{The energy change of the three lowest LL bands ($n = 0$ LL and $n = \pm 1$ LLs) at K valley versus Rashba SOC strength $\lambda_{R}$, based on Eq.~(\ref{EqB3}). 
The red, blue and black lines denote $n = 1$, $-1$ and $0$ LLs, respectively. We here set $\Delta_{so}=2\hbar eBv_{f}^{2}=1$.}
	\label{FIG10}
\end{figure}

\section{The effect of Rashba SOC}

In this appendix, we theoretically study how the topological phases are affected by Rashba SOC.
If the mirror symmetry is broken by a perpendicular electric field or by the interaction with a substrate, a Rashba SOC will arise in the Kane-Mele model with a term \cite{Kane1, Kane2}:
\begin{equation}
	H_{R}=\lambda_{R}(\sigma_{x}\tau_{z}s_{y}-\sigma_{y}s_{x})
\end{equation}
where $\lambda_{R}$ is the strength of Rashba SOC.
Without the real magnetic and pseudomagnetic field, the Rashba SOC has two effects on Kane-Mele model in Eq.~(\ref{Eq9}): (i) breaking the spin conservation along $z$ direction, (ii) destroying the intrinsic SOC gap 2$\Delta_{so}$. 
For $ 0 < \lambda_{R} < \Delta_{so} $, the energy gap 2 ($\Delta_{so} - \lambda_{R}$) remains finite and QSH effect keeps.  For $ \lambda_{R} > \Delta_{so} $, the energy gap closes and reopens with QSH phase disappearing \cite{Kane1,Kane2}.

When in the presence of both the Rashba SOC and the (pseudo)magnetic field $B$ ($B_{ps}$), the low-energy Hamiltonian for $K$ valley is:
\begin{equation}
	H_{K}=v_{f}[ \sigma_{x}\pi_{x}+\sigma_{y}\pi_{y} ]+\Delta_{so}\sigma_{z}s_{z}
	+\lambda_{R}(\sigma_{x}s_{y}-\sigma_{y}s_{x}).
	\label{EqB2}
\end{equation}
where $\vec{\pi}=\vec{p}+e\vec{A}$ or $\vec{\pi}=\vec{p}+e\vec{A}^{ps}$ . Under a real magnetic field, the Hamiltonian for $K'$ is related to $H_{K}$ by a time-reversal transformation:
$TH_{K}(p_{x},p_{y},\vec{A})T^{-1}=H_{K'}(-p_{x},-p_{y},-\vec{A})$. 
While under a pseudomagnetic field, the Hamiltonian for $K'$ is related 
to $H_{K}$ by a time-reversal transformation: $TH_{K}(p_{x},p_{y},\vec{A}^{ps})T^{-1}=H_{K'}(-p_{x},-p_{y},\vec{A}^{ps})$.
Taking $B$ as an example and solving the equation in Eq.~(\ref{EqB2}), the LL energy $E$ is ($q=0,1,2,...$): \cite{Martino}
\begin{equation}
    \left[\frac{E^{2}-\Delta_{so}^{2}}{2\hbar eBv_{f}^{2}}-(q+1)\right]\left[\frac{E^{2}-\Delta_{so}^{2}}{2\hbar eBv_{f}^{2}}-q\right]=4\lambda_{R}^{2}\left(\frac{E-\Delta_{so}}{2\hbar eBv_{f}^{2}}\right)^{2}.
	\label{EqB3}
\end{equation}
It can be verified this energy relation is both valid for $H_{K}$ and $H_{K'}$ no matter under a real magnetic or a pseudomagnetic field. 
It indicates the energy of LLs at $K$ and $K'$ valleys is similar, and their wavefunctions are related by time-reversal transformation. We can just pay attention on $K$ valley. The general solution of Eq.~(\ref{EqB3}) is a little more complex than results in Table.~\ref{Table1} and can be consulted in details in Ref.~\cite{Martino}. In Fig.~\ref{FIG10}, we focus on the energy variation of three lowest LLs ($n = 0$ LL
and $ n = \pm 1$ LLs) versus $\lambda_{R}$. For convenience, we take $\Delta_{so}=2\hbar eBv_{f}^{2}=1$. When $\lambda_{R}=0$, the $n = \pm 1$ LLs are spin-degenerate at $\pm \sqrt{2e \hbar B v_{f}^{2}+\Delta_{so}^{2}}$, while the zeroth LLs 
are spin degeneracy broken at $\pm \Delta_{so}$, which is consistent with the LLs distributions in Table.~I, Fig.~\ref{FIG3} and Fig.~\ref{FIG4}. 
As the $\lambda_{R}$ climbs, the spin along $z$ direction is no longer a good quantum number and the spin-degenerate $n = \pm 1$ LLs are split into two spin bands respectively (see Fig.~\ref{FIG10}). 
Differently, the $n = 0$ LL with higher energy stays at $\Delta_{so}$ always, 
while the $n = 0$ LL with lower energy starts to approach $\Delta_{so}$ from $-\Delta_{so}$ (no energy crossing). The former is attributed to $E=\Delta_{so}$ is always one solution of Eq.~(\ref{EqB3}) for $q=0$. 
Thus, the original SOC gap $2\Delta_{so}$ at $K$ and $K'$ valley within the zeroth LLs (denoted by black boxes in Fig.~\ref{FIG3} and Fig.~\ref{FIG4}) is shrunk by the Rashba SOC and QSH phase will be totally cannibalized once $\lambda_{R}$ is very large. 
Outside the SOC gap, since spin-degenerate LLs are lifted, the QH phases/QVH phases may evolve into spin-polarized QH phases/QVH phases.

\bibliography{ref}
\end{document}